\documentclass[sigconf]{acmart}
\acmConference[arXiv Preprint]{}
\acmYear{2026}

\usepackage[english]{babel}
\usepackage{blindtext}

\renewcommand\footnotetextcopyrightpermission[1]{} 
\setcopyright{none}

\usepackage{times}

\newcommand{\allnotes}[1]{}

\usepackage{graphicx}
\usepackage{times}
\usepackage{hyperref}
\usepackage{ulem}
\usepackage{xspace}
\usepackage{xcolor}
\usepackage{balance}
\usepackage{outlines}
\usepackage{tikz}
\usepackage{caption}
\usepackage{multirow}
\usepackage[]{url}
\usepackage{todonotes}
\usepackage{subcaption}
\usepackage{algpseudocode}
\usepackage{listings}
\usepackage{algorithm}
\usepackage{mfirstuc}
\usepackage{makecell}
\usepackage{xurl}

\author{
Sarah McClure$^1$, Tegan Wilson$^2$, Brad Karp$^{3,4}$, Michael Mitzenmacher$^5$, Sylvia Ratnasamy$^1$, Scott Shenker$^{1,6}$, and Minlan Yu$^5$
}
\affiliation{
$^1$UC Berkeley
$^2$Northeastern University
$^3$University College London
$^4$Google
$^5$Harvard University
$^6$ICSI 
}

\usepackage[subtle]{savetrees}
\usepackage{enumitem}

\usepackage[font={small},labelfont={small,bf},textfont={small,bf}]{caption}
\usepackage[font={small,bf},labelfont={small,bf},textfont={small},belowskip=3pt,aboveskip=2pt]{caption}
\usepackage[font={small,bf},labelfont={small,bf},textfont={small},belowskip=-1pt,aboveskip=1pt]{subcaption}
\usepackage[compact,small]{titlesec}

\usepackage{pifont}

\newcommand{\takeaway}[1]{\vspace{1pt}\noindent\fbox{\begin{minipage}{\columnwidth} \textbf{\textit{Takeaway: }} #1 \end{minipage}}\vspace{2pt}}

\newcommand{\ie}{\emph{i.e.,}\xspace}
\newcommand{\eg}{\emph{e.g.,}\xspace}
\newcommand{\etc}{\emph{etc.}\xspace}

\newcommand{\mypara}[1]{\noindent\textbf{#1}}

\renewcommand{\shortauthors}{McClure et al.}

\begin{document}
\sloppypar

\title{On Topology's Role in ML Training Performance}

\begin{abstract}
Modern machine learning training workloads run on large-scale networks of compute accelerators. 
The networks commonly deployed in these systems are typically variations of two basic topologies: the fat-tree Clos and the torus. In this paper, we derive analytical results the elucidate how the choice of topology shapes achievable performance for the small set of
collective communication operations that underlies modern machine learning workloads. We also consider how these results change when we include additional factors such as network failures and job placement strategies. Overall, we find that one topology does not dominate in all cases, but that the Clos achieves better collective completion time in most cases and provides benefits in resilience and flexibility.
\end{abstract}

\maketitle

\section{Introduction}\label{sec:intro}

Modern machine learning (ML) training workloads run on large-scale
ensembles of network-connected compute accelerators. At scales beyond
a single rack, built systems principally adopt two topologies: the
fat-tree Clos~\cite{alibaba, meta-rdma} and the torus (as exemplified
in the Tensor Processing Unit (TPU)~\cite{tpuv4}).\footnote{For
brevity, we refer to the former as Clos hereafter. Moreover, this paper focuses on ML training, not ML inference, which differs in some important characteristics.} For host
populations of equal size, these two topologies differ significantly
in bisection bandwidth, path diversity, and network
diameter -- differences one might expect to affect ML training
performance. A designer of a distributed ML training system must
choose an interconnect topology (or often, a hybrid of topologies, as
done in designs that employ one topology for ``scale-up,'' at limited
scale, and another topology for ``scale-out,'' at wider scale). How
should they choose? This choice should be grounded in a thorough
understanding of \textit{ML workloads' achievable performance on these
two candidate topologies, the Clos and torus, each considered in
isolation, as a function of system scale in accelerator count.}

Further key factors interact with topology to determine training
performance:
\begin{itemize}[leftmargin=*]
\item When \textit{links fail,} how does training performance on a topology
  degrade?
\item How rigidly does performance depend on the \textit{placement of
  computation} in a topology?
\item How much do \textit{enhanced switch capabilities,} such as
  multicast, improve performance in a topology?
\end{itemize}

We take an analytical approach to answering these questions. In the
interest of tractability, we limit our analysis to these central
network design choices of topology, scale, link bandwidth, and
recovery strategies under link failures. Additional constraints that
we do not address (\eg power consumption, hardware layout)
clearly influence design as well, so we do not see this work as the
final word on the role of network topology in ML training performance. On the
contrary, we see it as a necessary, analytically rigorous step
toward elucidating how these central choices interact to determine
achievable ML workload performance.

Data center designers have long embraced the Clos as a general-purpose
interconnect for a broad range of distributed applications, as it
offers flexible bandwidth provisioning (with full bisection bandwidth
or oversubscription) and resilience to link failures; both of these
properties are dividends of the Clos's abundant path
diversity~\cite{Clos-1953,clos-dc-alfares,jupiter-rising}. Further,
the Clos scales to connect large host populations using switches of
limited radix. GPUs have proliferated as ML accelerators situated
within hosts in data centers, and so have inherited the Clos for
scale-out connectivity.\footnote{Alternative topologies and link
technologies, such as NVlink and UAlink, have found use for
scale-up. While our primary focus in this work is the scale-out
setting, where Clos-derived topologies are the norm, we consider the
use of full mesh topologies and NVlink in scale-up in
\S\ref{sec:hybrid}.}

The torus topology, by contrast, first saw use in interconnects for
parallel machines that targeted high-performance computing (HPC) and
scientific computing workloads, which feature tightly synchronized,
closely coupled computations across processors~\cite{iWarp,TRC,MasPar-MP1,Cray-T3D,kary-ncube}. A toroidal
interconnect physically corresponds to the pattern of data motion
among processors induced by matrix multiplication, and other common
operations in these workloads. Distributed ML training computations
also exhibit this tightly synchronized nature. Their contemporary
implementations incorporate multiple distinct \textit{parallelisms},
each of which tends to propagate values along a linear, sequential
chain of data dependencies that may span processor boundaries (e.g., as in DeepSpeed's
3D Parallelism~\cite[Figure 2]{microsoft-deepspeed-3d}).
Multi-dimensional tori seem a natural match: one may ``lay out''
distinct parallelisms along each of a torus's dimensions. 

While Clos and torus topologies have been known and used for decades
in diverse distributed computations, to our knowledge there has been
no analytical comparison of their performance for ML
training that accounts for placement of distinct parallelisms in the topology
or how performance degrades under link failure---factors we show
are material to performance.
This gap in the literature may in part reflect the relative
balkanization of the data-center networking and HPC networking
communities, which for many years focused on workloads with quite
different characteristics. What has changed is, we think, the
ascendancy of ML training as a common workload of broad interest, both
for Clos-connected GPUs and torus-connected TPUs~\cite{tpuv4}.

We observe that ML training's reliance on a small set of
\textit{collective communication operations,} -- AllReduce, AllGather, AlltoAll -- each of which dictates a
very specific data dissemination goal among processors, offers a
convenient lens through which to evaluate the ML training performance
of a topology. Our approach is to derive completion times under a
simple message delivery model for these canonical ML collective
communication operations on Clos and torus topologies, in both full
link availability and link failure scenarios, across axes including
scale in nodes, link bandwidths, and torus dimensionality. We further
offer intuition for how these bounds arise---that is, how these two
topologies constrain the communication induced by ML collective
communication operations.
Lastly, we apply our performance models to modern LLM case studies to determine how each topology would perform on a particular training setup.

Our models do not suggest a ``dominant'' topology, such that all
collective communication operations generally perform better on one or
the other of the Clos or torus. However, our results suggest that unless AllGather is your bottleneck for performance, the Clos topology is likely the better choice.
\section{Setting}
\label{sec:background}
  
\begin{table}
{
\small

    \begin{tabular}{|l|m{3cm}|l|}
    \hline
    \textbf{Variable} & \textbf{Definition} & \textbf{Default Value} \\
    \hline
    $\alpha$ & per-link latency & 1 us\\
    \hline
    $\beta$ & link bandwidth & 400 Gbps \cite{tomahawk6, meta-rdma}\\
    \hline
    $\gamma$ & reduction time per byte & 1 ps \cite{a100} \\
    \hline
    $m$ & message size & 100 MB \cite{meta-ccl,training-comm} \\
    \hline
    $k_t$ & dimensionality of torus & 3 \cite{tpuv4} \\
    \hline
    $k_c$ & switch radix in clos & 128 \cite{tpuv4, tomahawk6} \\
    \hline
    $n$ & number of nodes participating in the collective  & - \\
    \hline
    $N$ & number of nodes in the topology  & - \\
    \hline
    $h$ & max hops between nodes in $N$ in a Clos  & $\begin{cases}
    2, & \text{if } n < k_c / 2 \\
    4, & \text{if } k_c/2 \leq n < k_c / 2 \\
    6, & \text{otherwise}
    \end{cases}$ \\
    \hline
    \end{tabular}
    \caption{Variables used to model the performance of each topology. Default values are based on relevant link technologies and model parameters.}
    \vspace{-1em}
    \label{tab:variables}
}
\end{table}

\mypara{Workload.}
In this paper, we compare the bounds
on collective completion time,
focusing  on three main collectives used in ML training: AllReduce (AR), AllGather (AG), and AlltoAll (ATA). These capture the communication used for Data Parallelism, Tensor Parallelism, and Mixture of Experts respectively~\cite{alibaba, colossalai-parallelisms, rail-only}.\footnote{Pipeline parallelism is also a common methodology for training, but its analysis is fairly simple as it consists of point-to-point communication. Thus, we focus on the other, more complex communication patterns.} 
We assume training jobs will use multiple forms of parallelism at once, but we begin by analyzing each collective separately (\S\ref{sec:baseline}) before generalizing to multiple on one network (\S\ref{sec:placement}). 

As mentioned in prior work \cite{forestcoll}, these collectives can be composed (or reversed) in order to implement others. 
Accordingly, we only analyze the three listed above: AR, AG, and ATA.
Further, we note that in some cases, splitting a given message into several smaller messages may allow for additional composition. For simplicity, we do not consider message-splitting.

A given collective may be implemented via many different algorithms or ``schedules'' -- \ie the set of flows (source-destination pairs) used to reach the desired end-state of the collective (\eg ring, one-shot, recursive doubling, \etc). 
There is a large body of work on determining the best schedule of flows to complete an overall collective on an arbitrary topology \cite{te-ccl, sccl, taccl, syccl}. These works generally aim to find efficient approximations of an NP-hard problem \cite{forestcoll}. 
Instead, in our work, we consider two particular and notably symmetric topologies. 
Further, collective algorithms on these particular topologies have been well-studied \cite{cctheory-chan07, topologyaware-sack12, mpicoll-pjesivac05, optimalbucket-jain10, alltoallmess-bruck94}. Thus, we build on this prior work where available and continue to exploit the regularity for additional conclusions. 

\mypara{Topologies.}
As mentioned in \S\ref{sec:intro}, we focus on two topologies: Clos and torus. In what follows, we use $N$ to denote the number of nodes in the topology as a whole, and $n$ to denote the number of nodes in a particular collective; when $n<N$ the collective is embedded into the larger topology.
For the Clos, we a traditional 3-tier $k_c$-ary fat tree as an initial baseline network topology.
By default, we set $k_c$ to a realistic value (128, see Table~\ref{tab:variables}), and for the collective choose subsets of size $n$ (the size of the collective group) of the overall $N$-node Clos to participate.
When taking a subset, we assume the best-possible locality (\eg all under one top-of-rack (ToR) switch, if less than 64 nodes).
In this topology, compute nodes are directly connected to a ToR switch, and switches do not have any special in-network capabilities unless specified (\eg multicast, copying data, reductions). 
For the torus, we assume a $k_t$-dimensional torus where each node is directly connected to its two neighbors in each dimension.\footnote{We assume the torus is equal size in each dimension. Thus, $k_t$ and the total number of nodes specifies the entire topology.} 
Unless stated otherwise, we assume a torus of size $n$, rather than a subset of some larger torus (\ie $n=N$).\footnote{In practice \cite{tpuv4}, slices of a larger torus network can be made into an individual torus using optical circuit switches. }
We assume a single host is at each node in the graph and the nodes can forward multi-hop traffic without involving compute cores.

For both topologies, we assume nodes to be the GPUs / TPUs that are performing the training compute. Thus, each accelerator will have its own access to the network. In the torus case, this is seen in the TPUv4 \cite{tpuv4} ICI links, and reflects servers with per-GPU NICS \cite{dgx-h100, dgx-a100, rubin-b100}.

\mypara{Performance.}
As in prior work, we parameterize these topologies with the alpha-beta (or Hockney) model~\cite{te-ccl, cctheory-chan07, mpicoll-pjesivac05,ccsurvey-wickramasinghe16}. 
Latency of a link is represented with $\alpha$, while the transmission time per bit (inverse link bandwidth) is $\beta$. 
We denote the computation time for reductions on a chunk of data as $\gamma$. 
This simplified model ignores queuing delay, allowing us to evaluate the impact of topology without having to model fine-grained transport layer choices for issues such as congestion control and pacing.

Accordingly, our main metric when comparing topologies is the collective completion time (CCT), depending primarily on $\alpha$, $\beta$, and $\gamma$. 
When relevant, we refer to terms that depend on $\alpha$ as the ``latency cost'', and terms that depend on $\beta$ will be called the ``bandwidth cost'' \cite{cctheory-chan07, sccl}. 
Table~\ref{tab:variables} summarizes the variables that we use to model performance. It also lists modern ``reasonable'' default values that we use to illustrate performance at specific operating points. 

\mypara{Scale-Up and Scale-Out Networks.}
Notably, these two topologies can be combined with others to form an overall topology with both a ``scale-up'' (\eg a high-bandwidth, highly-connected network between close-proximity nodes) and ``scale-out'' (\eg a datacenter network) network \cite{99flops}. 
Generally, the torus is used as a scale-up network while a Clos is used as a scale-out.
While we address this matter in detail in \S\ref{sec:hybrid}, we make two observations that support direct comparison, the focus of this work: (1) torus deployments generally scale to much larger numbers of nodes than other scale up networks (\eg intra-rack meshes \cite{nvlink}), therefore becoming comparable to a Clos; and (2) as we explore more in \S\ref{sec:placement}, a single collective will likely only span one of these two networks (\eg residing within one rack or between peer nodes in different racks).

We acknowledge that our work focuses on a specific selection of collectives, topologies, and parameters. We make these choices for simplicity. Evaluation of other options such as other Clos variants~\cite{ meta-rdma, alibaba, porter-multiplane, jupiter-evolving}, rail optimization~\cite{rail-only, nvidia-rail-optimized}, reconfiguration~\cite{tpuv4}, etc., is left as  future work, although we believe our framework would be useful for such an evaluation.
\section{Baseline Comparison}
\label{sec:baseline}

{\renewcommand{\arraystretch}{1.5}
\begin{table*}
    \begin{tabular}{|c|c|c|}
    \hline
        \textbf{Collective} & \textbf{Clos} & \textbf{Torus} \\
     \hline
         AllReduce & $\log_2(n)(h\alpha_c + \beta_c m + \gamma m)$~\cite{cc-mpich-03} & $\frac{k_tn^{1/k_t}}{2}(\alpha_t + \beta_t m) + log_2(n^{1/k})k\gamma m$~\cite{cctheory-chan07}$\dagger$ \\
    \hline
        AllGather & $h\alpha_c + \beta_c m (n-1)$~\cite{fat-tree-collectives}$\dagger$ & $\frac{k_tn^{1/k_t}}{2}\alpha_t + \beta_t m \frac{n-1}{2k_t}$ ~\cite{cctheory-chan07}$\dagger$ \\
    \hline
        AlltoAll & $h\alpha_c + \beta_c m (n-1)$~\cite{fat-tree-collectives}$\dagger$ & $\frac{k_tn^{1/k_t}}{2}\alpha_t + \beta_t m (n-1) \frac{n^{1/k_t}}{8}$~\cite{efficientalltoall-basu24}$\dagger$ \\
    \hline
    \end{tabular}
    \caption{Comparison of best completion times for known algorithms for each considered collective on torus and Clos. $\dagger$ denotes extension of the referenced work.}
    \label{tab:baseline-ub}
    \vspace{-1em}
\end{table*}

}

We start by providing formulas for the baseline performance of each collective on each topology.
Table \ref{tab:baseline-ub} provides the minimum  completion time of our considered collectives on each topology for the best-performing known  algorithms.
These results were derived assuming store-and-forward switches, optimal routing\footnote{We assume that flows are placed to avoid congestion as long as there is sufficient capacity.}, and 
that only one collective is running on the topology. We relax the last assumption in \S\ref{sec:placement}.
Our analysis in some cases repeats that of prior work, and in others extends prior work (entries denoted with a $\dagger$) ~\cite{cctheory-chan07, mpicoll-pjesivac05,cc-mpich-03, fat-tree-collectives}; for brevity, we provide descriptions of the algorithms and bounds, but do not present a full derivation (though some additional explanation is in Appendix~\ref{app:failure-derivations}). 

\begin{figure*}
    \centering
    \begin{minipage}[t]{0.36\linewidth}
        \centering
        \includegraphics[width=1\columnwidth]{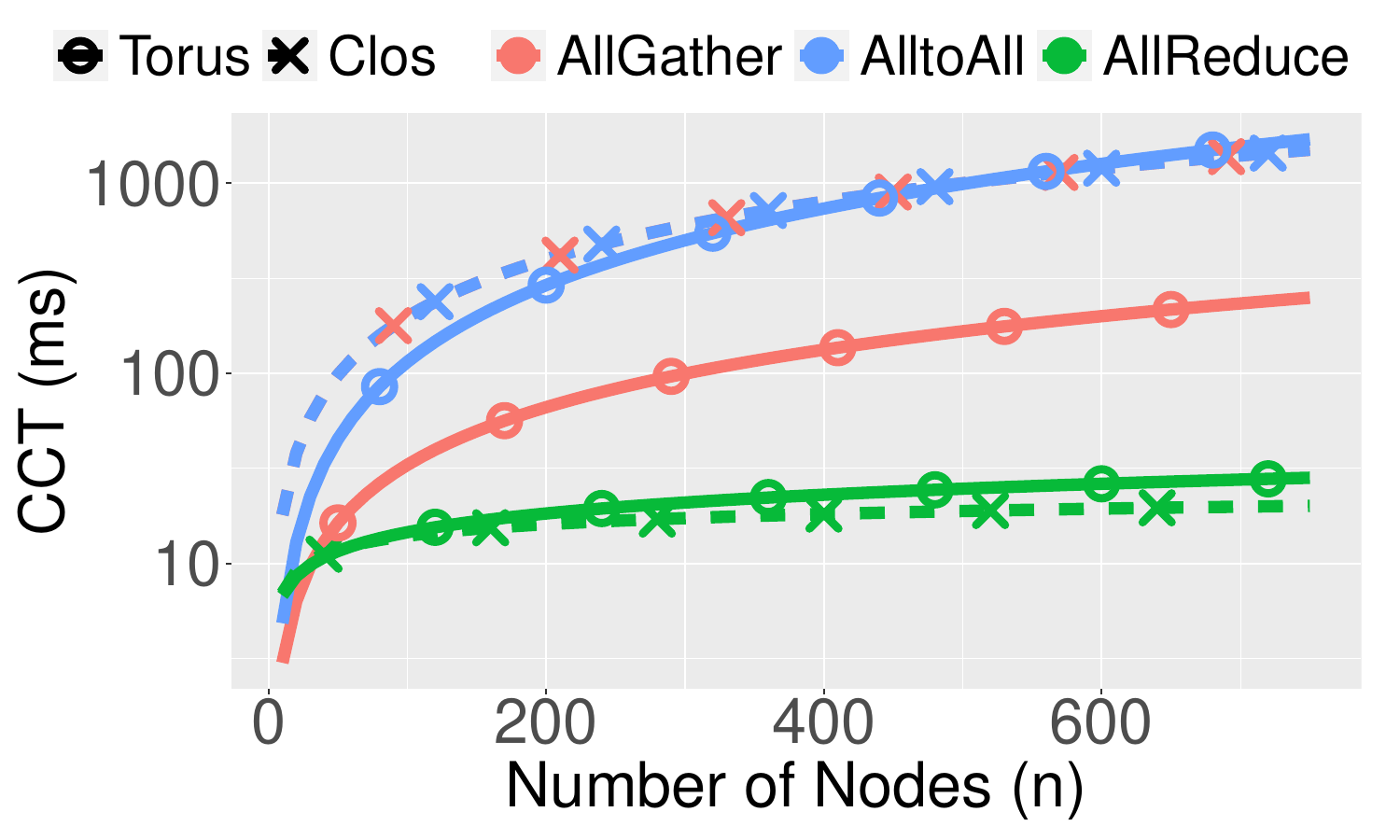}
        \caption{Completion times (using default values in Table~\ref{tab:variables}) of each collective on the torus and Clos as the number of nodes participating in the collective, $n$, scales.}
        \label{fig:baseline}
    \end{minipage} %
    \hspace{0.01\linewidth}
    \begin{minipage}[t]{0.34\linewidth}
        \centering
        \includegraphics[width=1\columnwidth]{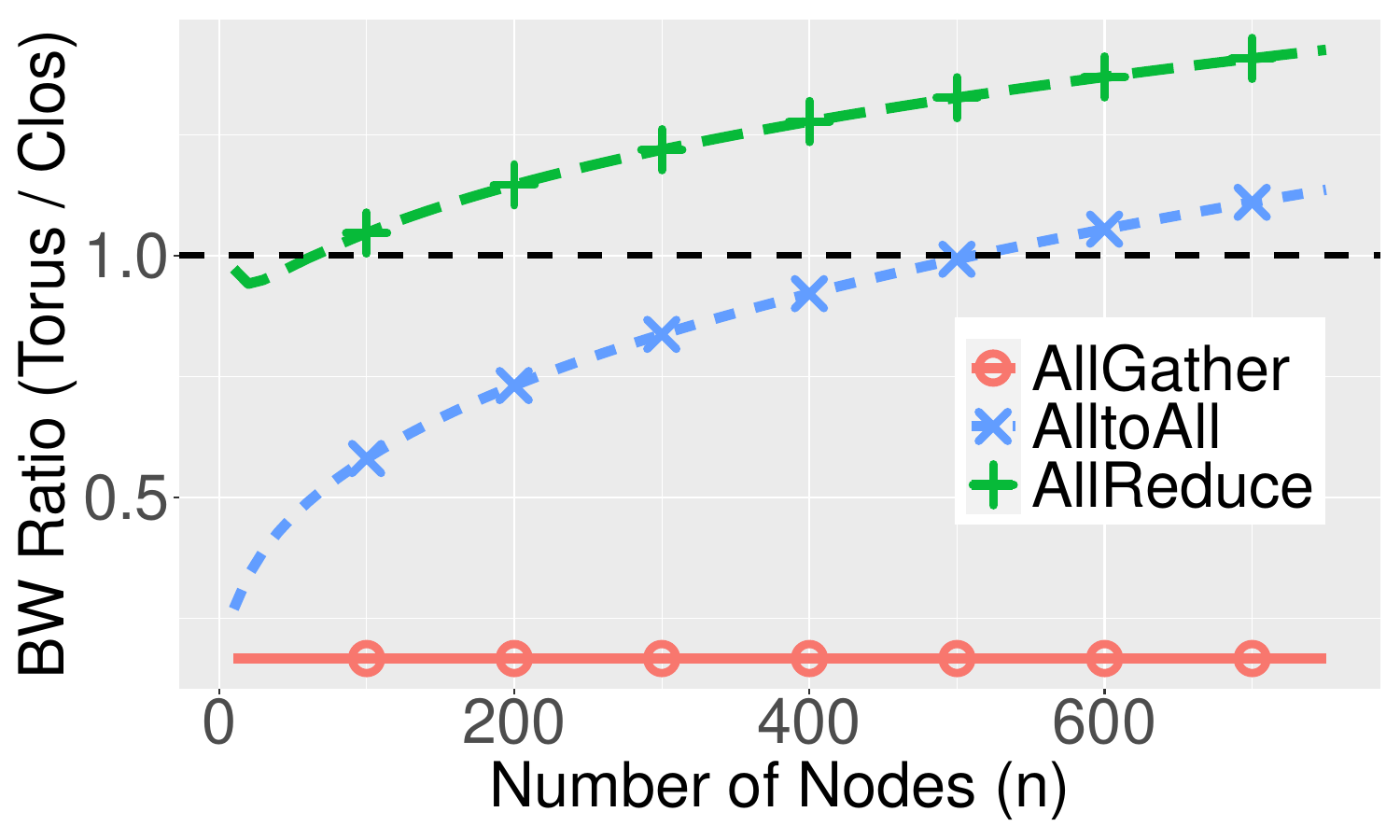}
        \caption{Ratio of bandwidth ($b_t/b_c = \beta_c / \beta_t$) required for the two topologies to achieve the same completion time for the default values given in Table~\ref{tab:variables}.}
        \label{fig:baseline-ratio}
    \end{minipage}%
    \hspace{0.02\linewidth}
    \begin{minipage}[t]{0.25\linewidth}
        \centering
        \includegraphics[width=\columnwidth]{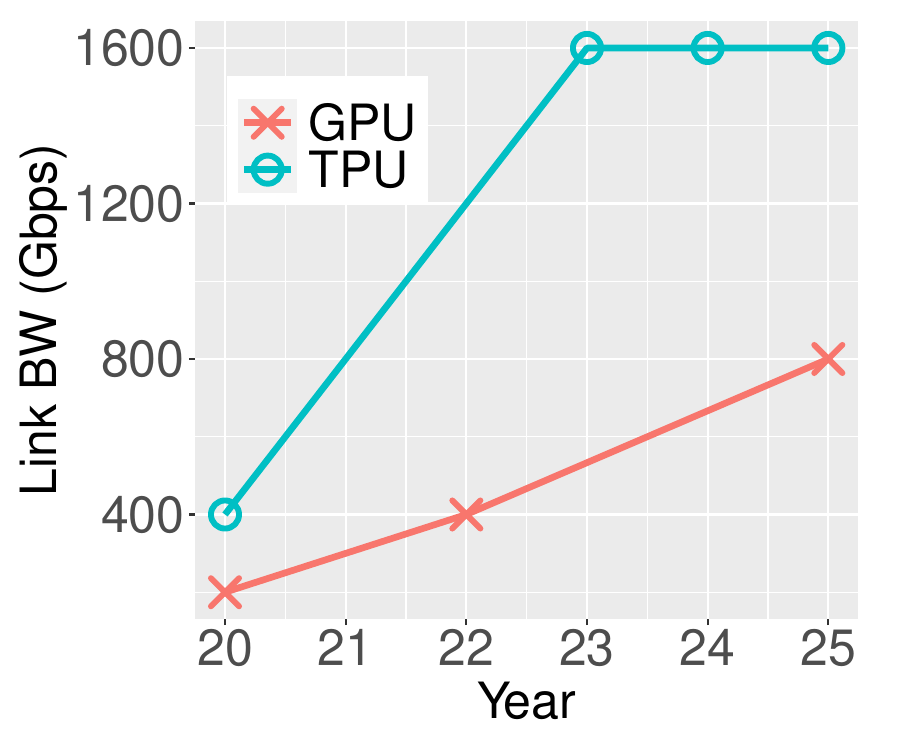}
        \caption{TPU and GPU per-link BWs over the last 5 years. \cite{dgx-a100, dgx-h100, rubin-b100, tpuv7, tpuv6e, tpuv5p}}
        \label{fig:gpu-v-tpu-bw}
    \end{minipage}
    \vspace{-1em}
\end{figure*}

We first provide a high-level overview of the results before discussing each result in detail.
For a more intuitive view, Figure~\ref{fig:baseline} plots the completion times (per  Table~\ref{tab:baseline-ub}) for the default values from Table~\ref{tab:variables} as $n$, the number of nodes participating in the collective, increases. 
For common deployment models today with multiple forms of parallelism, the number of nodes participating in a single collective together will likely be in smaller ranges of $n$ (\eg generally <100 in \cite{tpuv4}). 
In the figure, both topologies have the same basic parameters (link latency, message size, \etc as defined in Table~\ref{tab:variables}).

As shown in Figure~\ref{fig:baseline}, the better-performing topology depends on both the specific collective and the scale. 
Despite these nuances, there are some general patterns. 
At smaller scales, the torus often performs better or comparably to the Clos.
As $n$ grows, however, Clos eventually performs better in all collectives except AllGather (AG).
The torus's advantage in an AG is largely due to its additional access bandwidth. 
For all collectives, the torus performs well at small scales since nodes have high locality which keeps the diameter (which impacts all CCTs) low.\footnote{Though this result may be due to our assumption of maximally local placements; if at small scale the nodes are not placed to achieve locality, then the diameter of graph can increase significantly with $n$ and worsen the completion time.}
We discuss each collective in more depth later in this section. 

Notably, all of the plots show results for a 3-D torus. As we discuss later (\S\ref{sec:placement}), commonly multiple forms of parallelism will reside in one torus and be placed in different dimensions \cite{tpuv4-docs}. In this case, any one collective may only reside in one dimension. With $k_t = 1$, we note that the torus continues to outperform the Clos on AG, but it performs worse than the Clos at all scales as the diameter of the network is much larger and the number of links available for traffic is smaller compared to a higher-dimensional torus.

\takeaway{The topology that provides the lower completion time (between Clos and torus) generally depends on the number of nodes participating in the collective. However, torus is better in all cases for an AllGather, since the additional access bandwidth greatly affects the CCT. For a one-dimensional torus, the Clos outperforms in AlltoAll and AllReduce regardless of scale.}

\mypara{Varying Bandwidth.}
For a more general comparison (beyond a particular bandwidth normalization), we consider the relative bandwidth necessary for the two topologies to have the same completion time. 
Specifically, we find the ratio $r_b$ such that a Clos with per-link bandwidth $b_c$ has the same completion time as a torus with per-link bandwidth $b_t = r_b b_c$.  (Note $r_b$ is a function of $n$, $\alpha$, etc.) 
This analysis allows us to compare the scaling of the topologies without fixing their particular bandwidths. 
Torus and Clos link technologies or cost factors may allow bandwidths to scale differently over time (\eg TPUv4 \cite{tpuv4} mentions 400 Gbps links at a time when 100 Gbps links were the common upper bound for a traditional Clos-based datacenter \cite{jupiter-evolving}).
Thus, we seek to understand the potential of the topologies even as their link bandwidths differ.
Figure~\ref{fig:baseline-ratio} plots $r_b$ for each collective against the number of nodes in the topology. 
When $r_b<1$, Clos links need more bandwidth to match the torus on that collective, and vice versa when $r_b > 1$. 

To gain some insight from this graph, one can look at a particular value of $r_b$ and determine which topology will complete each collective faster at a particular scale (\ie choose a line on the y-axis and compare to each collective line at different scales). 
The Clos has the lower completion time for $r_b$ below the line for a given collective, and the torus has the lower CCT above the line.
To determine a realistic value of $r_b$, we plot recent per-link bandwidths for both TPU and GPU deployments in Figure~\ref{fig:gpu-v-tpu-bw}. 
Using the most recent values, we see that $r_b = 0.5$. 
Thus, looking at Figure~\ref{fig:baseline-ratio}, AllReduce would be faster on the Clos (since $r_b=0.5$ is \emph{below} the line at most scales for the two to have the same completion time), ATA would be faster on a torus for only small $n$ ( approximately <50), and an AllGather would always be faster on the torus.

\takeaway{With realistic bandwidth ratios, at larger scales the Clos has better performance on AlltoAll and AllReduce, but not AllGather.}

\mypara{Per-Collective Explanation.}
For an AllReduce, the cost is bound by how quickly a message (or a reduction including it) can reach all nodes, rather than the bandwidth cost.
The all-to-all connectivity of the Clos topology lends itself to a recursive doubling approach, allowing the collective to complete in $\log_2(n)$ transmissions \cite{recursivedoubling-refenacht16, topologyaware-sack12}.
On a torus, the collective's performance is bound by the diameter of the graph, and largely results in performing a 1-D AllReduce in each dimension.
Thus, as shown in Figure~\ref{fig:baseline} and Table~\ref{tab:baseline-ub}, the Clos scales asymptotically better in the size of the topology, both in latency and bandwidth cost. 
In Figure~\ref{fig:baseline-ratio}, $r_b$ increases past 1 as $n$ increases as bandwidth alone cannot compensate for the increasing diameter of the network ($\frac{k_tn^{1/k_t}}{2}$) and corresponding latency and compute costs.

In some cases,\footnote{When the message bandwidth $\beta m$ dominates, and when messages can be discretized and the AllReduce can be performed across $n$ parallel chunks.} it may be more efficient to perform a ReduceScatter followed by an AllGather, each on $\frac{m}{n}$-size messages.
ReduceScatter can be implemented using AlltoAll and then reducing the components at each node.
Because we analyze AllGather and AlltoAll throughout the rest of the paper, we do not separately consider this strategy for AllReduce. 

For an AllGather, as shown in Table~\ref{tab:baseline-ub} the bandwidth costs between the two topologies differ by a factor of $\frac{1}{2k_t}$ where $k_t$ is the dimensionality of the torus. 
In both cases, a node must receive $n-1$ messages of size $m$ on its access links.
Accordingly, the difference in bandwidth terms comes from the different access bandwidths available to a node in each topology ($2k_t$ links connect to a node in a torus, but only 1 in a Clos).
If the per-node access bandwidth
is set to be equal in both topologies, the bandwidth costs are the same. 
For the Clos, the total latency cost of an AllGather is bound by the maximum number of hops between two nodes (at most 6 in a 3-tier fat trees, regardless of the number of nodes), while the torus latency scales with the diameter of the graph.
While the latency costs differ, as in the case of Figure~\ref{fig:baseline}, it is not significant in comparison to the bandwidth cost. 
Thus, the line for AllGather in Figure~\ref{fig:baseline-ratio} is roughly at $r_b = \frac{1}{2k_t} = 1/6$.

For AlltoAll, the full bisection bandwidth of the Clos topology allows it to achieve asymptotically better bandwidth scaling by a factor of $\frac{n^{1/k_t}}{8}$, and asymptotically better latency scaling ($h\alpha$ vs. $\frac{k_tn^{1/k_t}}{2}\alpha$).
However for a small enough network size, the constant latency scaling of the Clos is worse than torus, leading us to the relationship we see in Figure~\ref{fig:baseline}.

\takeaway{On both topologies, AllToAll and AllGather are more bandwidth-bound than latency-bound, though AllGather specifically is bound by \emph{access} bandwidth, not overall network bandwidth. AllReduce, instead, is bound by latency. These properties give rise to the results seen previous in the section.}

\begin{figure}
    \begin{minipage}{0.9\columnwidth}
            \includegraphics[width=\columnwidth]{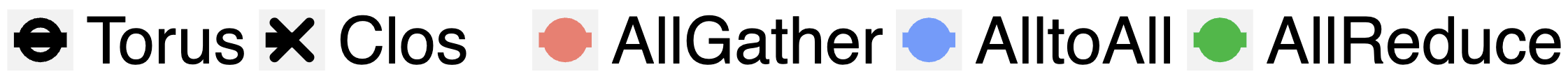}
    \end{minipage}\\
    \begin{minipage}[t]{0.48\columnwidth}
          \centering
          \includegraphics[width=\columnwidth]{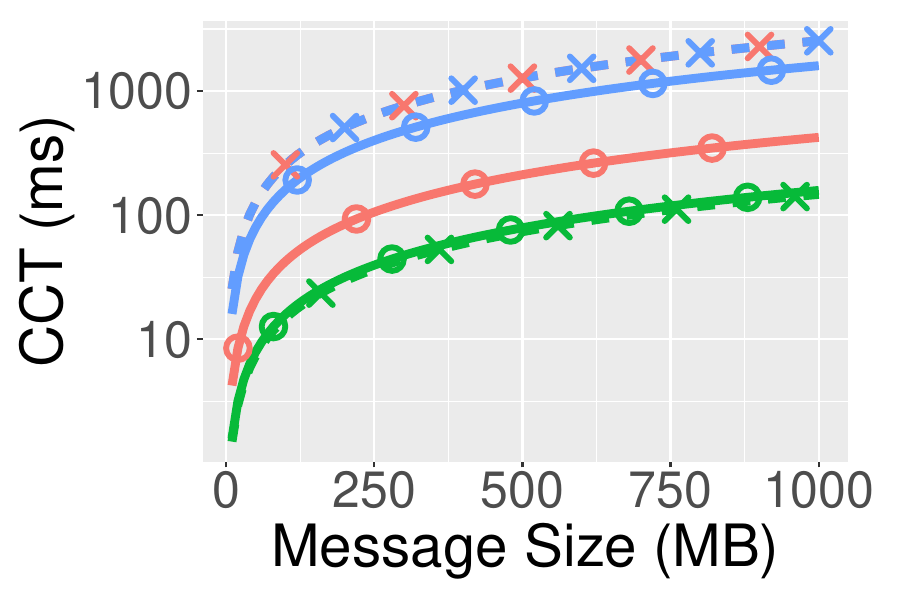}
        \caption{Completion times for each collective on a 128 node network as the message size is varied.}
        \label{fig:message-size-sweep}
    \end{minipage}%
    \hspace{0.02\linewidth}
    \begin{minipage}[t]{0.48\columnwidth}
        \includegraphics[width=\columnwidth]{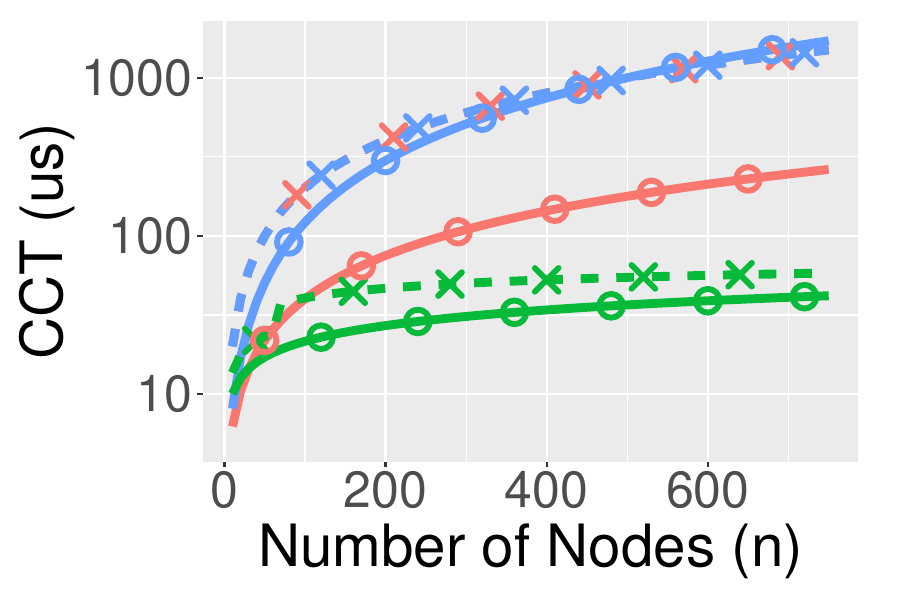}
        \caption{Absolute completion times of each collective on the torus and Clos with \textbf{very small message sizes (100KB)}.}
        \label{fig:baseline-smallm}
    \end{minipage}
    \vspace{-2em}
\end{figure}

\mypara{Different Message Sizes.}
Lastly, we present a similar analysis, but now vary the message size, $m$, and keeping the number of nodes constant at $n=128$ (a typical collective group size in \cite{meta-rdma, tpuv4}).
As shown in Figure~\ref{fig:message-size-sweep}, the message size (at least in these ranges) has no effect on the relative order of the two topologies.
We only found a small difference with extremely small message sizes (less than 500KB).
In this scenario, the number of hops between nodes becomes very important to the CCT, especially when there are dependencies between flows.  
Accordingly, the torus achieves a much lower CCT for AR in this case as the diameter of the torus is smaller than $h\cdot log(n)$ (the scaling of the latency term in the Clos). 
Notably, if the number of nodes participating in the AR is smaller than the total number of nodes in the topology (as would be typical), this effect is smaller as the max distance between two nodes could be smaller (\eg for nodes in one rack, the scaling factor would be $2\cdot \log_2(n)$\footnote{In this case, implementing AR as an AG and a local reduction would produce better CCT.}). 
Otherwise, the relative performance of the two topologies on AllGather and AlltoAll remain mostly the same.

\takeaway{As message size varies, the relative performance of the torus and the Clos remain mostly the same. Only very small message sizes change the ordering of the topologies for an AllReduce.}

Perhaps unsurprisingly, Figure~\ref{fig:baseline} and closer inspection of the equations in Table~\ref{tab:baseline-ub} demonstrate that neither topology is better in all cases. 
The best performing topology depends on the collective and assumptions about the relative values of $\alpha$, $\beta$, the message size, the number of nodes, etc. 
For the values of these constants in Table~\ref{tab:variables}, we find that the Clos topology generally does better for very large numbers of nodes, but the torus does about the same or better for small groups.
With all of these variables in mind, we now consider many potential considerations a network operator may have to consider such as failure  and how the workload is placed on the network in order to refine our understanding on if there is, in fact, a topology that consistently has better performance. 

\section{Failures}
\label{sec:failures}

{\renewcommand{\arraystretch}{1.75}
\begin{table*}
{
\small
    \begin{tabular}{|c|c|c|c|c|c|c|}
    \hline 
        \multirow{2}{*}{\textbf{Topology}}  & \multirow{2}{*}{\textbf{Link Type}} & \multicolumn{2}{c|}{\textbf{AllReduce}} & \multicolumn{2}{c|}{\textbf{AllGather}} & \textbf{AlltoAll} \\
    \cline{3-7}
        & & \textit{Net-level} & \textit{Sched-level} & \textit{Net-level} & \textit{Sched-level} & \textit{Net-level}  \\
    \hline
        \multirow{2}{*}{Clos} & L1 & 
        $\log_2(\frac{N}{R}) \beta_c m$ & 
        $2\alpha_c + \beta_c m$ &
        $\left( \left\lceil \frac{R(N - R)}{R -1} \right\rceil \! - \! (N \! - \! 1) \right) \beta_c m$&
        $\frac{N-1}{2}6\alpha_c$  &
        $\left( \left\lceil \frac{R(N - R)}{R -1} \right\rceil \! - \! (N \! - \! 1) \right) \beta_c m$
        \\
    \cline{2-7}
          & L2 & 
          $\log_2(\frac{N}{P}) \beta_c m$  &
          $2\alpha_c + \beta_c m$  & 
          $\left( \left\lceil \frac{P(N - P)}{P -1} \right\rceil \! - \! (N \! - \! 1) \right) \beta_c m$ &
          $\frac{N-1}{2}6\alpha_c$ &
          $\left( \left\lceil \frac{P(N - P)}{P -1} \right\rceil \! - \! (N \! - \! 1) \right) \beta_c m$
          \\
    \hline
        Torus  & - &
        $2\alpha_t + \beta_t m$  &
        $\alpha_t+\beta m$  &
        $2\alpha_t + \frac{N-1}{2k_t} \beta_t m$ &
        $\frac{(N-1)}{2k_t(2k_t-1)} \beta_t m$ \ddag &
        $\frac{(N-1)N^{1/k}(k-1)}{8(kN-1)}$ \ddag
        \\
    \hline

    \end{tabular}
    \caption{Additional time added to the completion time under a single link failure for each topology. For brevity, we define $R = k_c/2$ (the number of nodes in a rack in a Clos) and $P = k_c^2/4$ (the number of nodes in a pod). More complete derivations are included in Appendix~\ref{app:failure-derivations}. Entries with \ddag~are a lower bound rather than based on a known routing/schedule solution. }
    \label{tab:failure}
    \vspace{-1em}
}
\end{table*}
}

\begin{figure*}
    \begin{minipage}[]{0.69\linewidth}
        \centering
        \begin{subfigure}[]{0.5\linewidth}
            \includegraphics[width=\columnwidth]{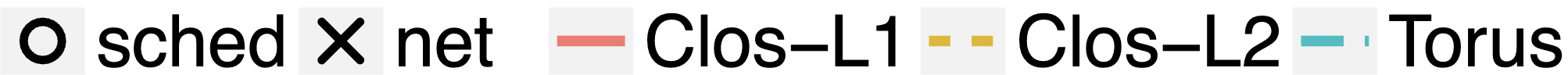}
        \end{subfigure}\\
        \begin{subfigure}[]{0.32\linewidth}
            \includegraphics[width=\columnwidth]{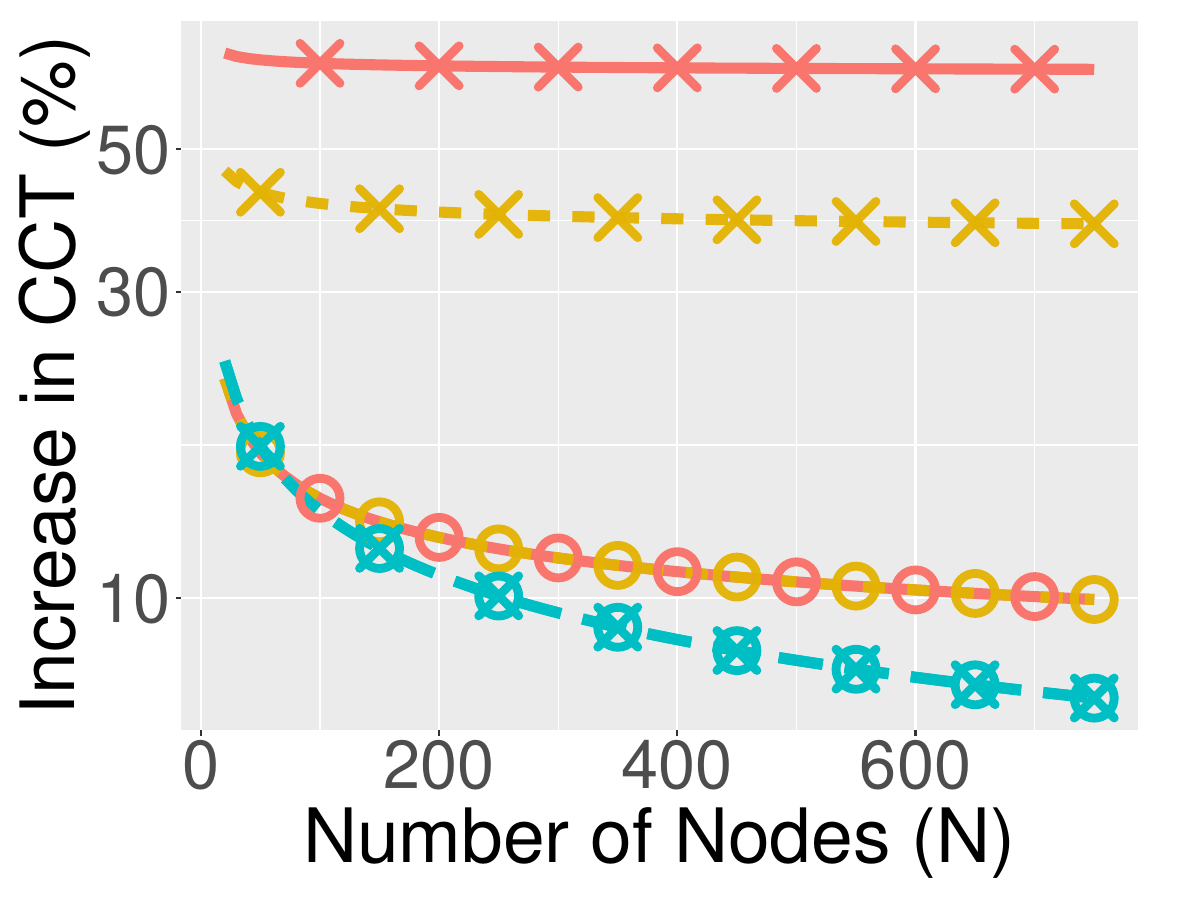}
            \caption{AllReduce }
            \label{fig:fail-ar}
        \end{subfigure}%
        \begin{subfigure}[]{0.32\linewidth}
            \includegraphics[width=\columnwidth,trim={0 0 0 0cm},clip]{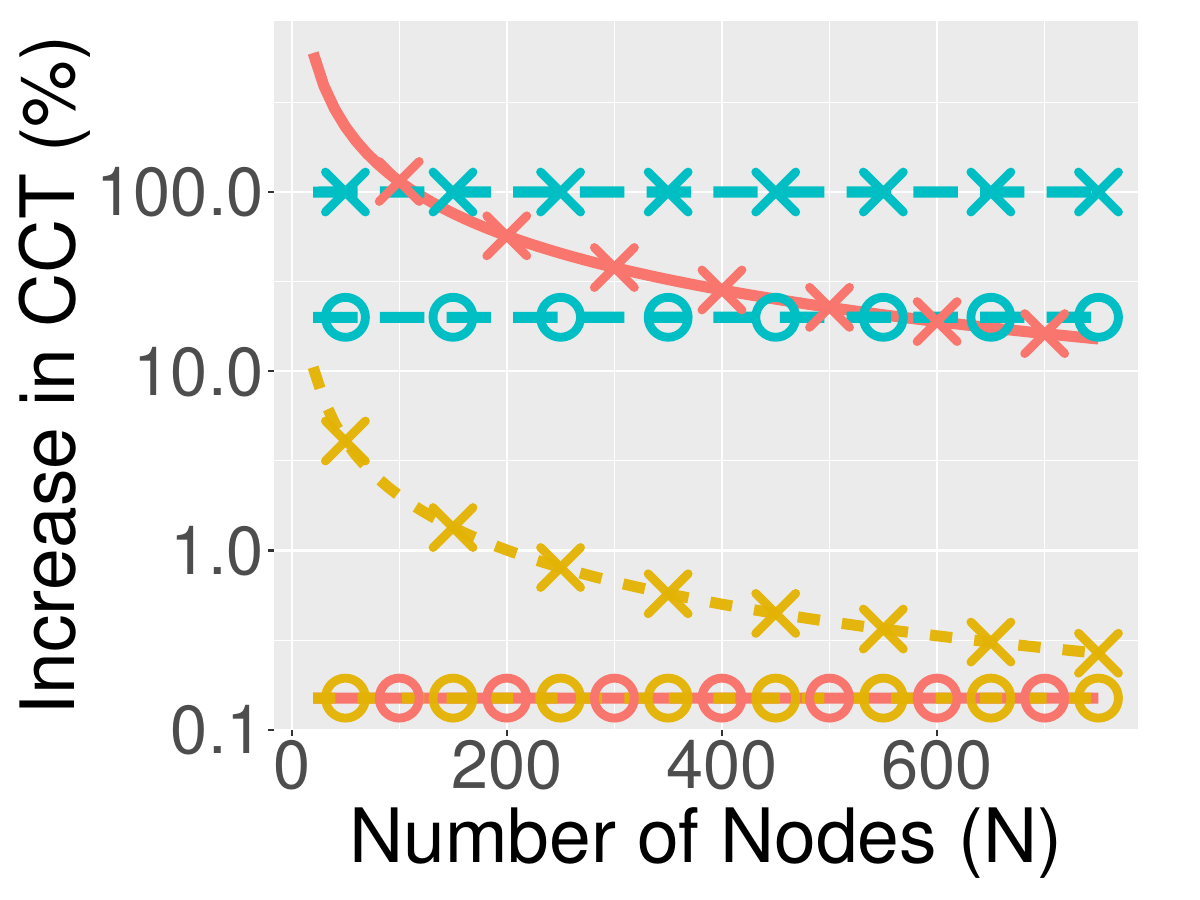}
            \caption{AllGather}
            \label{fig:fail-ag} 
        \end{subfigure}%
        \begin{subfigure}[]{0.32\linewidth}
            \includegraphics[width=\columnwidth,trim={0 0 0 0cm},clip]{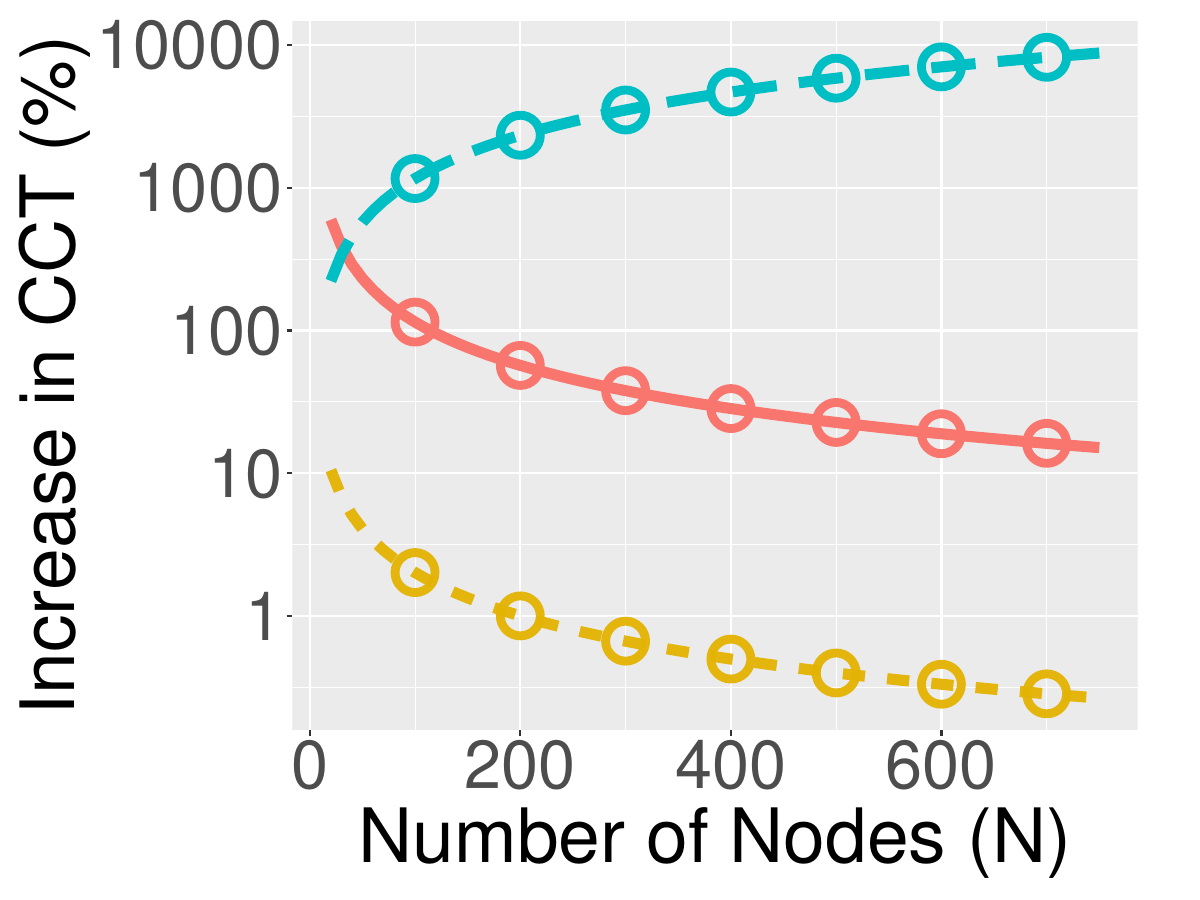}
            \caption{AlltoAll}
            \label{fig:fail-ata} 
        \end{subfigure}
        \caption{Impact on completion time (normalized to baseline) for each topology, failure type, and mitigation approach. Values are drawn from the defaults in Table~\ref{tab:variables}.}
        \label{fig:failure}
    \end{minipage}%
    \hspace{0.02\linewidth}%
    \begin{minipage}[]{0.29\linewidth}
        \centering
        \includegraphics[width=0.7\columnwidth]{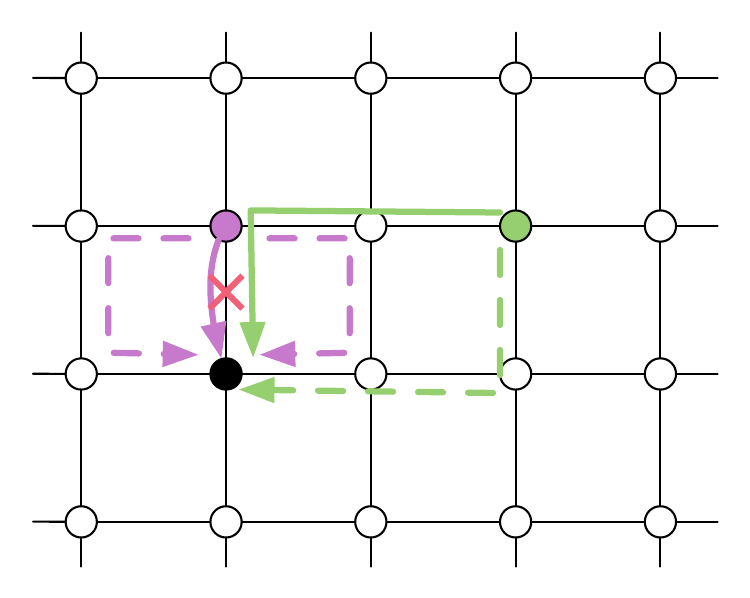}
        \caption{Torus link failure alternate routes example for a single-hop path and a multi-hop path.  Alternate paths of equal or best-possible length are shown (dashed).}
        \label{fig:torus-failure}
    \end{minipage}
    \vspace{-1em}
\end{figure*}

When operating a network at scale, link failures are inevitable \cite{tpu-resilience, msft-failures, meta-reliability, meta-failures-imc}.
Though more complicated patterns can occur, for simplicity we model link failures as single, bidirectional, failures for the entire duration of a given collective, 
Thus, we seek to determine how well each of these topologies can handle such network failures -- \ie how much the possible collective completion time changes when there is a non-partitioning failure. 
One topology may have more natural resilience to such failures, especially since this is a feature for which traditional Clos-based datacenter networks are explicitly designed~\cite{clos-dc-alfares, vl2}. 

In the torus, since all links are equivalent, we pick the failed links at random. 
For Clos, we separately consider when links between different tiers of the network fail. 
Specifically, we separately consider ``L1'' links between ToRs and aggregation switches and ``L2'' links between aggregation and core switches. 
We do not consider downlinks from ToRs as these would cause a network partition and a collective could not complete.\footnote{As mentioned in \S\ref{sec:intro}, the scale-up network connecting the nodes may be used to avoid a partition in such a case, but we ignore the implications of the additional network for this analysis.}
Importantly, here we do \emph{not} fix $k_c$ for the Clos and instead scale it to the minimum size necessary to fit $N$ (\ie $n=N$). This ensures there is not a large portion of the network (and bandwidth) that is unused and could be used to help in the case of a failure. Thus, moving forward, \emph{we assume $n=N$ for both topologies}.

Much of the extensive literature on fault-tolerance for MPI collectives focuses on node failures/partitions~\cite{ftmpi, mpich-v, stellner1996cocheck, sankaran2005lam} or transport-layer mechanisms for handling network congestion due to failures~\cite{mpi-networkfailures}. 
In \cite{tpu-resilience} and \cite{alltoall-torus-failure}, practical routing mechanisms for handling faults in a torus are discussed.\footnote{Google's solution to link failures in a torus is described in \cite{tpu-resilience}. The approach considers multiple, potentially non-shortest alternate paths that take a ``wild'' hop in a different dimension than the failure and uses an ILP that optimizes for ATA throughput (with additional constraints) to choose between candidate paths. This approach enables these routes to be calculated offline and installed into simple switches.}
Instead of a specific algorithm, we assume that traffic will be placed on a next-shortest path with the least impact on the given collective's CCT -- and given the symmetry of the network and workloads, any shortest path has the same impact. 
Rather than imposing a particular mechanism, we aim to determine the fundamental, best-case impact on completion time under single link failures.

\mypara{Network-Level Mitigation.}
First, we consider what \emph{network-level} approaches -- such as changes to routing and load-balancing -- can do to avoid the failure under the established schedule of flows given by the collective algorithm. In network-level mitigations, traffic matrices on the network \emph{do not change}; only the path between a prescribed source and destination can change.

A summary of our results is shown in Table~\ref{tab:failure} (``Net-level`` columns) with the \textit{additional} time necessary shown for each topology and mitigation strategy. 
Derivations are included in Appendix~\ref{app:failure-derivations}, though we provide some intuition behind the results in this section.
Note that we introduce two new variables for brevity, $R=k_c/2$ and $P=k_c^2/4$, which represent the number of nodes in a rack and a pod in a Clos respectively.
These results are visualized in Figure~\ref{fig:failure} to easily compare across mitigation strategies 
(point shapes) and across topologies and link failure types (colors and dash lengths). 
In the figure, the additional time is normalized to the original completion time of that topology and collective found in \S\ref{sec:baseline}.

For network-level mitigations on a Clos, the network must re-route flows using the failed link to one of the other shortest paths (of which there are many in a Clos). 
Thus, this increases the bandwidth cost proportionally, up to the number of affected flows, as they must share links with other flows. 
In AllGather and AlltoAll, these displaced flows  can be spread evenly among the remaining paths (\eg $N-R$ flows on $R - 1$ links for L1). 
Accordingly, as $N$ increases and the flow an individual link becomes smaller relative to the overall traffic matrix, so does the impact on CCT. In AllReduce, however, there is only one flow per host and path in each step of the schedule, so a full extra message of bandwidth cost must be incurred multiple times (\eg $\log_2(\frac{N}{R})$ for L1).

In a torus, the optimal schedules for both AllGather and AllReduce use no multi-hop flows (as every node needs the same data). 
Thus, the next-shortest path in the event of a link failure requires bypassing the failed link using three hops through a different dimension of the torus as shown in Figure~\ref{fig:torus-failure}.\footnote{This assumes a torus with $k_t > 1$ and $N^{1/k} \geq 4$. We do not consider tori outside these constraints as they have different next-shortest paths.}
Since we assume all nodes are participating in the collective, this will generally result in oversubscription of the links used to avoid the failure.\footnote{However, in practice, when multiple parallelism are used on one network, other links may not be in use at a particular time. While we discuss the issue of placing multiple parallelisms on one torus in \S\ref{sec:placement}, we do not consider this case for failures.}

As shown in the table, much like a Clos network, the impact on the completion time with network-level mitigations is a function of the number of flows that must be re-routed ( $\frac{N-1}{2k_t}$ for AG).
As the next-shortest paths are 2 hops longer than the one-hop paths in the base schedule, this adds latency cost as well as a bandwidth cost that is linear in $N$. 
In an AllReduce, the CCT is determined by the total path delay between two points in a ring thus it increases by a constant latency and bandwidth cost with the now-longer and congested path.
In AlltoAll at the network level, almost every path is multi-hop and can be re-routed onto many alternate shortest paths. 
Thus, additional latency cost is not incurred. 
However, there is an increase in bandwidth cost as the total capacity of the network decreases. 

\takeaway{With network-level mitigation of a link failure, the Clos has lower impact on CCT than the torus for AllGather and AlltoAll due to its highly-redundant bandwidth. In AllReduce, a particular link failure only increases the latency around a single ring  and thus the torus has lower impact on CCT.}

\mypara{Schedule-Level Mitigation.}
Next, we consider another class of mitigations: \emph{schedule-level} (also in Table~\ref{tab:failure} and Figure~\ref{fig:failure}). Here, we assume the schedule (\ie series of traffic matrices) determined by the collective algorithm/library \emph{can} be changed based on the failure. Meanwhile, we will continue to assume optimal routing underneath. Notably, this would require some mechanism to notify the collective library (\eg NCCL \cite{nccl}) about the link failure and change the schedule accordingly. Here, we seek to determine the potential impact of such a system, and do not consider how it might be implemented.

At the schedule-level for AllGather, not much can be done while following the class of schedules that achieve the lowest completion times as there are no dependencies between flows.
Instead, we assume the application strategically uses a schedule which does not require all links in the network (at the cost of dependencies between flows). 
The associated latency cost with introducing these dependencies is shown in the table.
For AllReduce, only a constant overhead is necessary at the schedule level as a node can be excluded for most of the algorithm and receive the final reduction at the end.

In AllReduce on the torus, the 1-D ARs can be ordered to only require an additional transmission between neighboring nodes.
For AllGather on the torus, the completion time must increase as it is bound by the access bandwidth of the nodes (and two nodes will have lower access bandwidth due to the failure). 
In both topologies, nothing additional can be done at the schedule level for AlltoAll as there are no dependencies between flows to manipulate.

\takeaway{With schedule-level mitigation, both topologies can improve their CCT over network-level. As with network-level, the Clos can handle failures better for AlltoAll and AllGather, but the torus is better for AllReduce.}

Overall, schedule-level mitigation for a torus is generally better since modifying the schedule to avoid a link can leverage unused bandwidth, while naively using the next-shortest paths with routing / load balancing results in paths conflicting with other flows.
For Clos, there are many schedules with the same bandwidth cost that the application may use to implement the collectives, but these create different tradeoffs in resilience and latency. 
Clos networks are able to leverage their redundant paths to efficiently re-route the flow(s) while the torus must rely on schedule-level mitigations to avoid overhead from routing around the failure.

We note that this analysis was limited to single failures. We leave extensions to arbitrary numbers of failures to future work.

\section{Placement} \label{sec:placement}

\begin{table*}
    \begin{tabular}{|c|c|c|}
    \hline
        \textbf{Placement} & \textbf{AlltoAll and AllReduce} & \textbf{AlltoAll and AllGather} \\
    \hline
         Typical & $\alpha \sqrt{N} + \beta m \left(\frac{N}{8} + \frac{\sqrt{N}}{2}\right) + \frac{1}{2}\gamma m \log_2\left(N\right)$ &  $\sqrt{N}\alpha + \beta m \left( \frac{\sqrt{N}}{2} + \frac{N}{8} \right)$ \\
    \hline
        Locality-Opt. & $2 \alpha\left( \sqrt{N} + N^{1/4} \right) + \beta m \left(N^{1/4} + \frac{1}{4} N^{3/4}\right) + \frac{1}{2}\gamma m \log_2(N)$ & $2 \alpha\left( \sqrt{N} + N^{1/4} \right) + \beta \frac{m}{4} \left( \sqrt{N} + N^{3/4} \right)$ \\
    \hline
    \end{tabular}
    \caption{Sum completion time of two collectives on a 2-D torus for typical placement (each collective operates on a single dimension of the torus) versus the locality-optimized placement optimizing for AlltoAll at the cost of the other collective. 
    }
    \label{tab:placements}
    \vspace{-1em}
\end{table*}

\begin{figure}
    \centering
    \begin{minipage}[t]{0.38\columnwidth}
    \centering
        \includegraphics[width=\columnwidth]{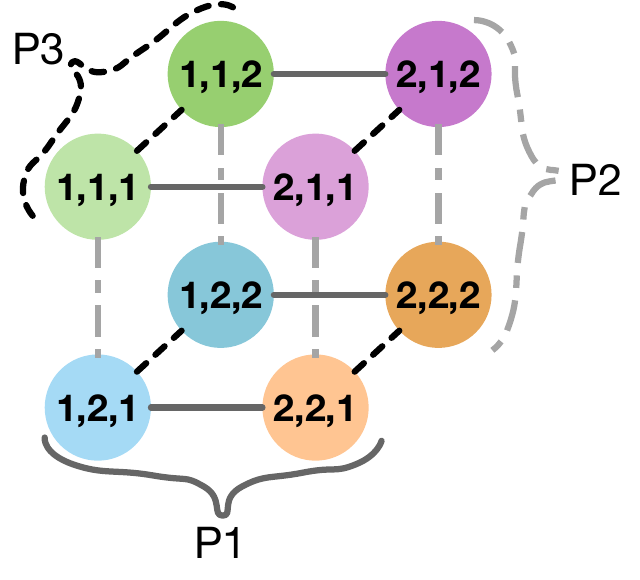}
        \caption*{Logical structure of a training job with 3 forms of parallelism (P1-P3). }
    \end{minipage} 
    \hspace{0.01\columnwidth}
    \begin{minipage}[t]{0.59\columnwidth}
    \centering
       \includegraphics[width=0.9\columnwidth,trim={0 0 16cm 0}, clip]{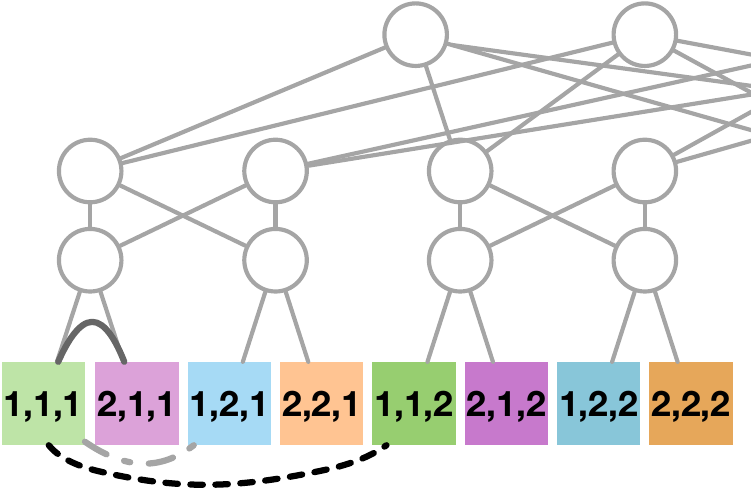}
        \caption*{Training job placed in (half of) a Clos network. Collective groups for each parallelism are shown for node 1,1,1.}
    \end{minipage}
    \caption{A training job has a logical structure due to its parallelism (and their sizes) which must be embedded/placed into the chosen physical topology. This determines which nodes perform which collectives together in the physical network.}
    \vspace{-1em}
    \label{fig:placement-intro}
\end{figure}

Thus far, we assumed that all nodes participate in the same single collective.
However, this is not how modern machine learning workloads are deployed. 
Workloads often use multiple forms of parallelism, parallelized across different axes~\cite{tpuv4, topo-opt, colossalai-parallelisms}. 
Instead of a single communication collective among all nodes in the network, nodes participate in smaller collectives across (possibly multiple) orthogonal axes, as shown in Figure~\ref{fig:placement-intro}.
This logical structure must then be embedded or ``placed'' into the physical topology.\footnote{Job placements in tori is a well-explored problem, but often under the assumption of \emph{many independent jobs} with the goal of optimizing for locality of nodes \cite{slurm-placement-torus, chen2025coadaptingmachinelearningjob}.}
This placement defines which nodes participate in which collectives, and thus impacts how traffic will be traversing the network. An example placement on a Clos network is shown in Figure~\ref{fig:placement-intro} for a small job with three forms of parallelism.
Accordingly, for a particular topology, the set of well-performing placements may be flexible or rigid (\ie there can be one optimal placement, or many equally-good placements).
We compare the flexibility of placement decisions on Clos versus torus, and how these decisions affect communication collective performance.

In a non-oversubscribed Clos, there is no significant difference between placements. 
Bandwidth cost does not change, and latency cost can increase by at most an additive $4\alpha_c$, the maximum latency difference between any two node pairs.
In a torus, however, placement may change both latency cost and bandwidth cost. 
All collectives require latency cost at least $\alpha_t$ times the maximum distance between any two group members.
In addition, AlltoAll workloads require bandwidth cost at least $\frac{\beta_t (n-1)}{2k_t}$ times the average distance between group members, where $n$ is the size of the group.

The typical placement in a torus is to devote each dimension to a different form of parallelism (resulting in groups of size $N^{1/k_t}$ placed along each axis) \cite{tpuv4-docs}.
Each collective then operates on the ring of connections between group members, resulting in a maximum distance of $\frac{1}{2}N^{1/k_t}$ and an average distance of $\frac{1}{4}N^{1/k_t}$ between group members.
Many worse placements exist.
For example, placing all communicating nodes randomly results in a factor $k_t$ increase in both maximum and average distance, directly resulting in a factor $k_t$ increase in latency cost for every workload, and an additional factor $k_t$ increase in bandwidth cost for AlltoAll workloads.

\begin{figure}
    \centering
    \begin{minipage}[t]{0.45\columnwidth}
    \centering
        \includegraphics[width=0.9\columnwidth]{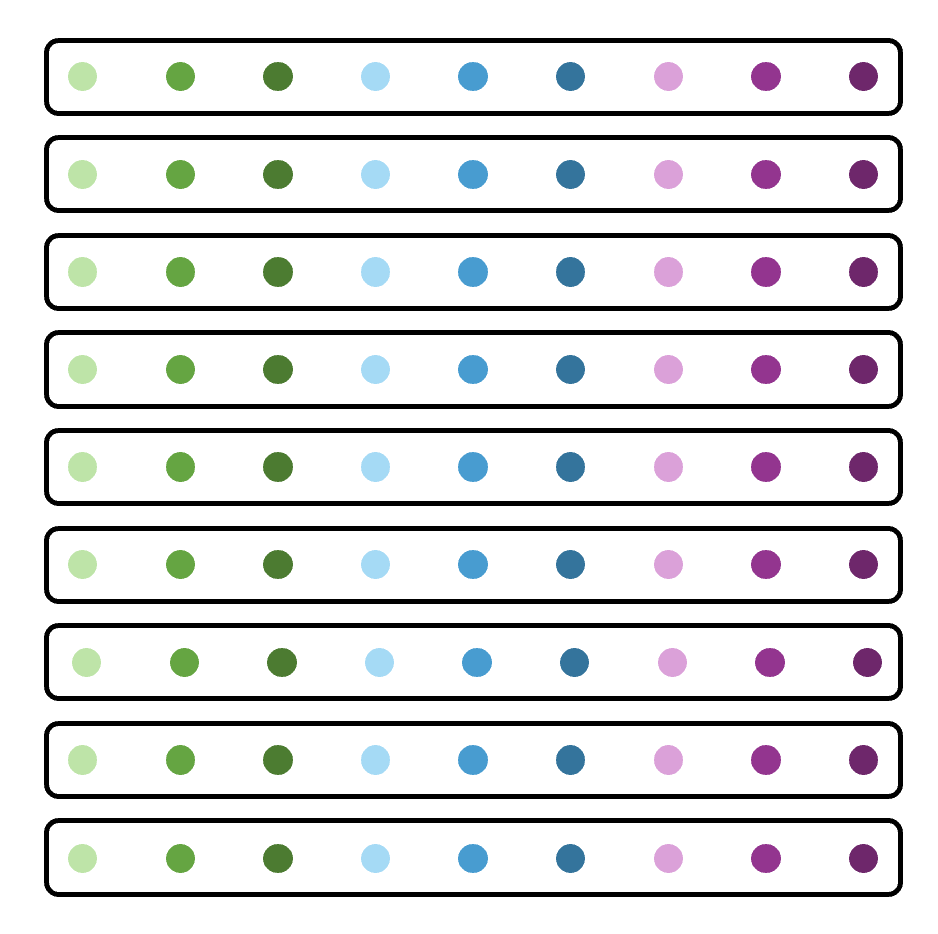}
        \caption*{Typical 2-D torus placement}
        \label{fig:trad-torus-placement}
    \end{minipage} 
    \hspace{0.05\columnwidth}
    \begin{minipage}[t]{0.45\columnwidth}
    \centering
       \includegraphics[width=0.9\columnwidth]{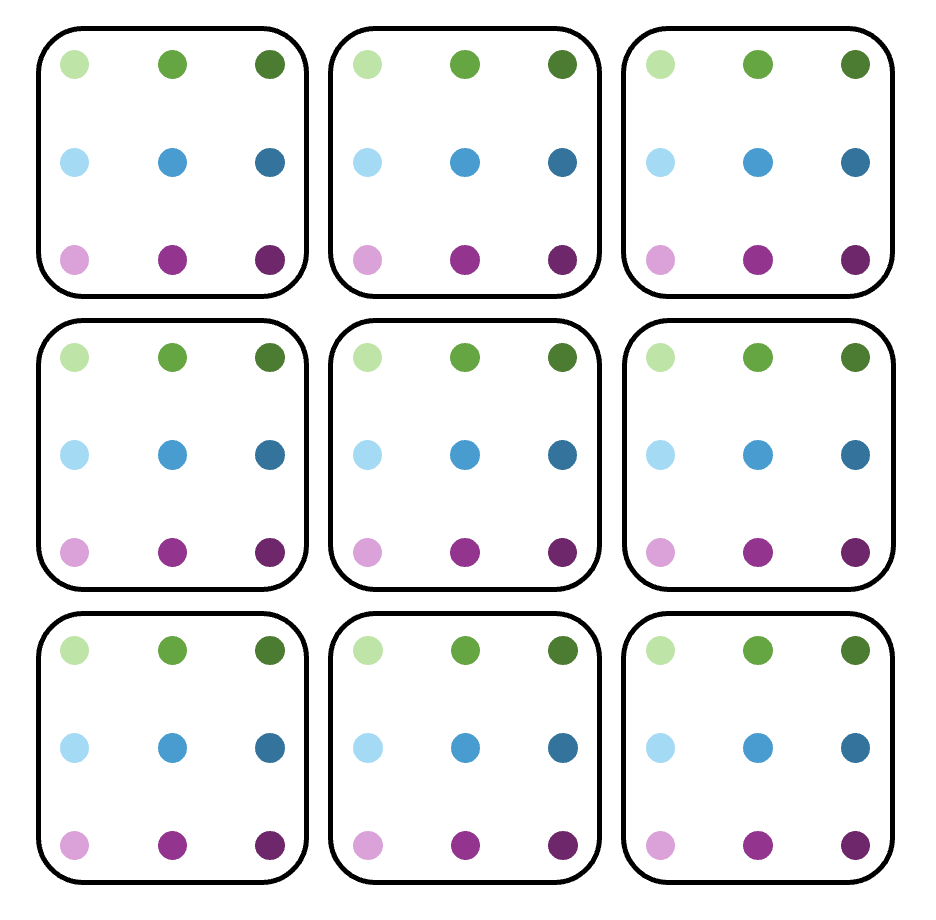}
        \caption*{Locality-optimized placement optimized for one parallelism}
        \label{fig:alt-torus-placement} 
    \end{minipage}
    \caption{Placement of two parallelisms on a 2-D, $9\times 9$ torus. One parallelism is nodes of the same color, the other is nodes that are boxed together.}
    \vspace{-2em}
    \label{fig:placement-visual}
\end{figure}

Optimizing for locality in one parallelism will improve its performance, but will inevitably hurt other parallelisms due to their orthogonal nature. 
However, it is worth investigating when this tradeoff may be beneficial.
For example, consider the 2-D symmetric torus. 
For a given parallelism, one can optimize for short distances between group members by selecting groups which tile the torus with $\sqrt{N}$ smaller 2-D grids\footnote{These grids are like 2-D tori, only they do not have wraparound links.}, as in Figure~\ref{fig:placement-visual}.
This results in a maximum distance of $2N^{1/4}$ between group members (and an average distance of approximately half that), an asymptotic improvement over the typical placement.
However, for the second parallelism, each group must contain one member from each of the previous groups, requiring twice the distance between members than in typical torus placement.

If we consider the total additive cost, then when both parallelisms are running the same collective type (\eg both running AllReduce), the locality-optimized placement is clearly sub-optimal, as it increases the second collective's cost by a factor of two. 
However, it asymptotically improves over the typical torus placement when the optimized parallelism is AlltoAll, as shown Table~\ref{tab:placements}.
This is because unlike other collectives, distance between group members affects both the latency and bandwidth cost of AlltoAll workloads. 
Of course, this assumes that the two parallelisms run at non-overlapping times and have equal message sizes. 
We confirm this finding and explore cases with different message sizes by applying the locality-optimized placement to real models in \S\ref{sec:case-studies}.

\takeaway{While placement doesn't affect Clos performance, it affects torus greatly, and different placements are better for different collective types and parallelizations. The torus provides opportunity for tuning improvements due to placement, but is not flexible as arbitrary placements can cause serious performance problems that do not exist for the Clos.}

\section{Cost}

Another significant difference between these two topologies for a network operator to consider is cost. In order to investigate this difference, we consider many topological factors that may impact cost, rather than determining the exact cost of implementing one of these topologies today. Thus, we consider three main factors: ports per switch, number of switches, and total link bandwidth in the network (\ie number of links times the link bandwidth). The values of each of these parameters are shown for both the torus and the Clos in Table~\ref{tab:costcomponents}.

{\renewcommand{\arraystretch}{1.75}
\begin{table}
    \begin{tabular}{|c|c|c|}
    \hline
        \textbf{Parameter} & \textbf{Torus} & \textbf{Clos} \\
        \hline
        Ports/switch & $2k_t$ & $k_c$ \\
        \hline
        Switches & $N$ & $\frac{5}{4}k_c^2$ \\
        \hline
        Total BW & $Nk_t \cdot \frac{1}{\beta}$ & $3N \cdot \frac{1}{\beta}$\\
        \hline
    \end{tabular}
    \caption{Various metrics affecting cost for each topology.}
    \label{tab:costcomponents}
    \vspace{-1em}
\end{table}
}

\begin{figure}
    \centering
    \begin{subfigure}[t]{0.47\columnwidth}
        \centering
        \includegraphics[width=\columnwidth]{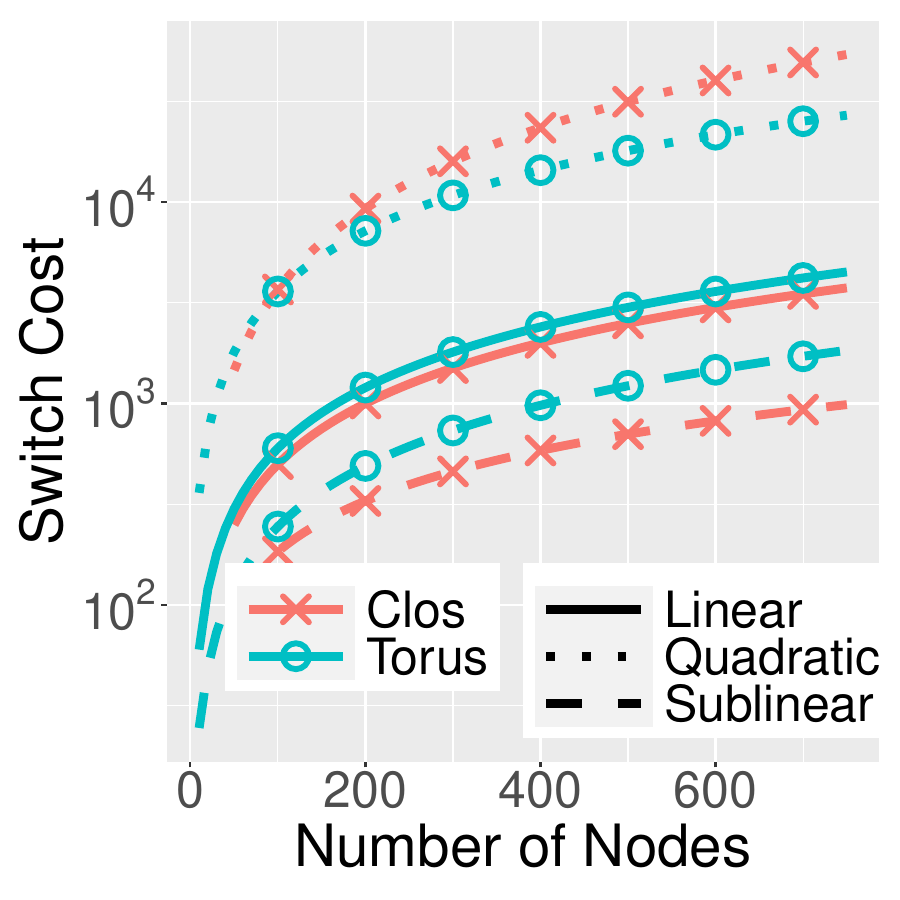}
        \caption{Total switch cost of each topology across sizes for different per-switch pricing schemes.}
        \label{fig:cost-schemes}
    \end{subfigure}%
    \hspace{0.02\columnwidth}
    \begin{subfigure}[t]{0.47\columnwidth}
        \centering
        \includegraphics[width=\columnwidth]{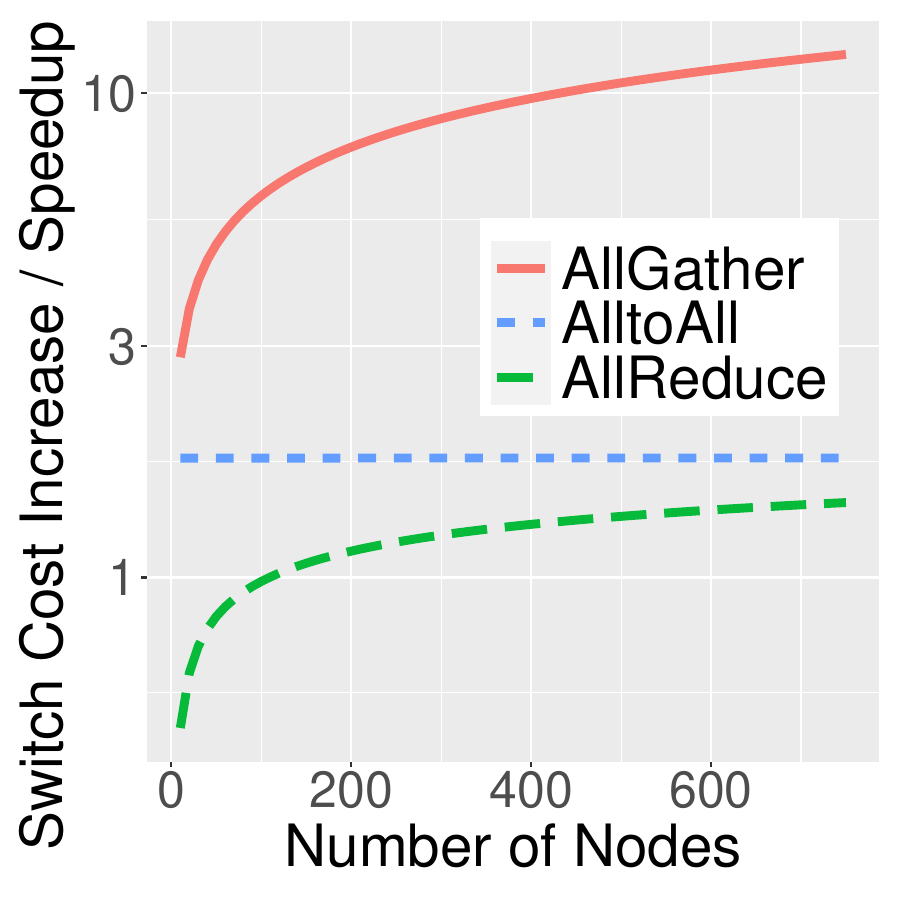}
        \caption{Clos switch cost increase (for quadratic pricing scheme) for unit speedup over torus.}
        \label{fig:cost-for-speedup}
    \end{subfigure}
    \caption{Switch cost comparisons of the two topologies.}
     \vspace{-1em}
\end{figure}

The torus will generally have switches with much smaller port counts as $k_c$ is a function of the network size in a Clos (\eg 32 for a roughly 8,000 node network), while $k_t$ is the dimensionality of the torus (generally 2-3 \cite{tpuv4, tpuv5p, tpuv6e, tpuv7}). Meanwhile, as seen in the table, the number of switches in the Clos is lower than that of the torus (since $k_c^3 /4 = N$, $\frac{5}{4}k_c^2$ < $N$ for $k_c > 5$). 
Lastly, the total bandwidths of the two networks are quite similar. A 3-dimensional torus and a Clos have an equal number of links (and therefore total bandwidth). This is because a torus has $k_t$ unique links per node, while a 3-tier Clos has $N$ links at each of 3 tiers.

Given these baselines, in Figure~\ref{fig:cost-schemes}, the total switch cost of each topology is shown for three possible basic cost functions: switch costs that are linear, quadratic, and sublinear in the number of ports per switch. We do not choose these functions for a realistic model of switch cost (which will likely be more complex, depending on area and power \cite{router-area, router-cost}), but rather to determine the general relationship between the costs of the topologies. 
As shown in the figure, Clos is cheaper with linear and sublinear cost functions, but more expensive with quadratic cost.

Even in the quadratic case when the Clos is more expensive, the additional cost may be worth the speedup that comes with the Clos. 
Accordingly, we define two relative measures: the price ratio, $p = Price(Clos) / Price(Torus)$, and speedup for each collective, $s_c = CCT_c(Torus) / CCT_c(Clos)$ (taken from \S\ref{sec:baseline}). For example, $p=2$ means the Clos costs twice as much as the torus and $s_c = 2$ means the Clos completes the collective $c$ twice as fast as the torus.
We plot $p/{s_c}$ for each collective in Figure~\ref{fig:cost-for-speedup} to compare the increase in cost to the performance benefit between the two topologies.
As shown in the figure, the relative cost of the Clos over the torus for the speedup gained depends on the collective, and becomes more expensive with size for AllGather and AllReduce. 

We note that while we do consider the total bandwidth of the topologies, we leave a complete analysis of cabling cost to future work as this will depend on the layout and other constraints, but we do not expect this to be a dominant factor in the cost as long as all links are optical \cite{datacentercost, helios}. Instead, our analysis provides a few first-order approximations of the relative costs of each topology.

\takeaway{The most significant difference in cost parameters between the two topologies is port count per switch. Unless the cost per switch is super-linear in the number of ports, the Clos has a lower overall cost even with the higher port counts under our pricing models.}
\vspace{-0.5em}

\section{Hybrid Topologies}
\label{sec:hybrid}
Our analysis thus far has ignored the potential of hybrid topologies that include different networks for the scale up and scale out networks. 
We now compare two common deployments with scale-up and scale-out networks: intra-racks meshes in a Clos scale-out network and tori connected via a Clos-based datacenter network.

In deployments that do not use a torus, the scale-up network is usually an extremely high-speed and well-connected network such as NVLink \cite{nvlink}. While these networks can be very high performance, their scale is inherently limited due to the topologies used. For example, the GB200 NVL72 \cite{nvlink-details} is a full mesh between all GPUs in the rack. In fact, there are 18 parallel meshes between the 18 server trays in the rack (with additional communication links intra-server). While such a topology would be desirable for any network, a full mesh has obvious scaling problems. Thus, these topologies are reserved for racks and servers, and do not apply to entire datacenter networks. Accordingly, these GPUs will also be connected to a typical datacenter fat tree network. 
For this analysis, we will generalize the NVLink topologies to redundant meshes ($N$ full meshes for $N$ GPUs) and refer to these topologies as simply a ``mesh''.

In torus deployments, the torus is also limited in its scale \cite{tpuv4, tpuv7}. Thus, each server is also connected to a typical datacenter network. For the purposes of this analysis, we will assume that each compute node has its own access to the datacenter network (\ie the NICs in \cite{tpuv4, scaling-book} are statically partitioned for each TPU in the tray).

We now compare the performance possible on these hybrid topologies. Effectively, we ask, \textit{how does having access to these additional networks change the potential performance?} 
Notably, in many deployments, these two networks will be used for different dimensions of parallelism (\eg tensor parallelism within a rack, then data parallelism between racks) \cite{topo-opt, torus-multislice, scalingmegatron}. For these cases, our previous analysis applies (though does not include an intra-rack mesh by itself) and we use it for these cases in \S\ref{sec:case-studies}. Thus, we instead consider how these hybrid topologies may be used in \emph{conjunction} for one collective between multiple nodes both in the same scale-up and scale-out networks.

In general, we assume that the scale-up network will have higher bandwidth \cite{torus-multislice, nvlink}; we denote its $\beta$ as $\beta_u$ and assume it is smaller than the scale-out (Clos) network's $\beta_o$. 
We also assume this bandwidth difference is substantial enough that paths through the scale-up network are preferred.
The results for a Clos + mesh hybrid can be seen in Table~\ref{tab:hybrid}. Note that we again use $P$ and $R$ for the number of nodes in a pod and rack respectively.
All of these results build off of the CCTs found in \S\ref{sec:baseline}.

{\renewcommand{\arraystretch}{1.75}
\begin{table*}
{\small
    \begin{tabular}{|c|c|c|}
        \hline
        \textbf{Collective} & \textbf{Clos + Mesh} & \textbf{Clos + Torus} \\
        \hline
        AllReduce & $\alpha + \beta_u m + \gamma mR + log_2(\frac{N}{R})(6\alpha + \beta_o m + \gamma m)$ & 
        $\frac{k_tP^{1/k_t}}{2}(\alpha + \beta_u m + \gamma m) + log_2(\frac{N}{P})(6\alpha+\beta_o m + \gamma m)$\\
        \hline
        AllGather & $6\alpha + \beta_o m(\frac{k_c^2}{2}-1) + \alpha + \beta_u m(\frac{k_c^2}{2}-1)$ &  
        $ 6\alpha + \beta_o m(k_c-1)) + \frac{k_tP^{1/k_t}}{2}\alpha + \beta_u m(k_c-1) \frac{P-1}{2k_t}$\\
        \hline
        AlltoAll &  $6\alpha + \beta_o m (N-R)$ & $6\alpha + \beta_o m(N-P) $ \\
        \hline
    \end{tabular}
    }
    \caption{Collective completion time (CCT) on hybrid topologies. Additional explanation and derivations are provided in Appendix~\ref{app:hybrid-deriv}. }
    \label{tab:hybrid}
    \vspace{-1em}
\end{table*}

}

First, we consider a Clos with an intra-rack mesh.
In the case of an AlltoAll, the intra-rack links can be used for intra-rack flows, decreasing the number of messages that must traverse the scale up network.
For AllReduce, the additional connectivity can be exploited to do intra-rack reductions ($\alpha + \beta_u m + \gamma mR$) before reducing across racks via the Clos (last term).
Similarly, in an AllGather, nodes can exchange their messages with their counterparts in other racks via the Clos while performing an intra-rack AG. Once inter-rack transmissions have completed $\left( 6\alpha + \beta_o m \left(k_c^2/2-1\right) \right)$, each rack can perform another intra-rack AG with the new messages $\left( \frac{k_tP^{1/k_t}}{2}\alpha + \beta_u m(k_c-1) \frac{P-1}{2k_t} \right)$.

\begin{figure}
    \centering
    \includegraphics[width=0.85\columnwidth]{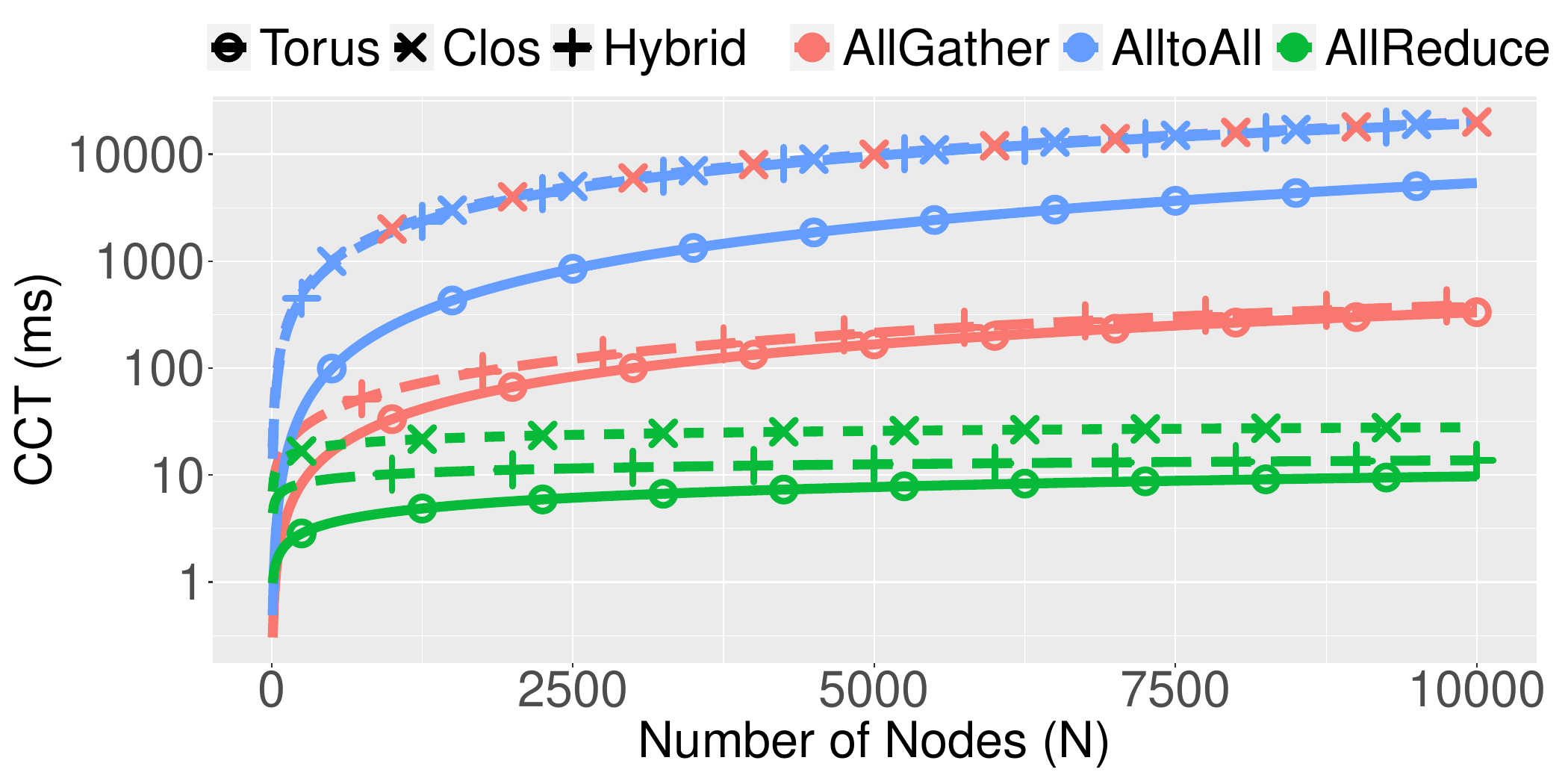}
    \caption{Comparison of Clos + Torus hybrid topology with total nodes N with a single torus  with $\beta_u$ and a Clos with $\beta_o = 10\beta_u$ \cite{dgx-h100} both of size N. The hybrid topology performs better than just a scale-out network and can approach the performance of just a scale-up, high-bandwidth torus for some collectives (\eg AR).  }
    \label{fig:hybrid}
\vspace{-2em}
\end{figure}

In the torus case, we consider a Clos network with $N$ nodes, where each \textit{pod} is connected internally via a torus interconnect (reflective of \cite{tpuv4}). Similarly to the mesh version, the scale-up network can provide some performance improvements. While the scale-up network is not as well-connected as a mesh, it does span beyond a single rack. 
The results for the torus follow a similar pattern to the mesh.
For an ATA, a normal ATA is conducted in the torus, while inter-pod flows must travel the Clos and thus bound the CCT.
For AR, a local reduction is done within the pods before reducing between pods. 
Lastly, for an AG, an AG is done within the pods while nodes exchange the necessary messages across pods, then perform another AG within the pods with the new data. 

The results for the torus + Clos hybrid are also shown in Figure~\ref{fig:hybrid} to demonstrate the benefit of a hybrid topology. For reference, we compare to just having the Clos network (with lower bandwidth) and a large torus (size $N$) with no datacenter interconnect. Overall, we see that the AlltoAll does not benefit much from the hybrid topology (most flows are still between pods), but AllReduce and AllGather do benefit substantially.

\takeaway{Hybrid topologies can offer benefits over just a Clos, though this performance is best observed when using a mesh or when the bandwidth in the the scale-up network is higher than that of the Clos. }
\vspace{-0.75em}

\section{Multicast}

We now consider one example in a broader category of switch capabilities: multicast. Generally, the choice of topology may impact how much switch capabilities can improve the performance and resilience of a collective. Here, we seek to determine how multicast, a common switch capability, can change these properties. Considering that collectives such as AllGather effectively requires each node to perform a broadcast, multicast may save a significant amount of transmission cost.

First, we consider a Clos topology performing an AllGather. We assume our baseline optimal schedule in which every node sends messages to all other nodes simultaneously. Multicast offers two benefits in this scenario: fewer link transmissions are necessary to complete the collective and accordingly, the number of unused links at any given point increases (\ie more flexibility to route around failures). 
We find that  $\frac{1}{4}(5k_c^3 - 4k_c^2 - 8k_c -12)$ transmissions are saved per node by using multicast compared to the unicast baseline schedule (derived in Appendix~\ref{app:multicast}).
Further, in the baseline schedule, there are times when every link is utilized. In the multicast case, capacity is available throughout the schedule for other traffic (\eg control messaging) or to tolerate link failures.
While these transmissions are saved, the completion time does not change as it is bound by the time for each node to receive $n-1$ other messages and the highest latency between two nodes.

Other collectives do not benefit as directly from multicast. An AlltoAll, for example, has nothing to gain from multicast as each flow between all pairs of nodes is a unique piece of data. An AllReduce, similarly does not have any broadcasts in our baseline implementation. However, if implemented as an AllGather followed by a Reduce (or a series of such operations), the same benefits as the AllGather follow.

For a torus, multicast can reduce the bandwidth cost associated with distributing a chunk of data as dependencies between flows can be relaxed. For example, a node sending a message to along one row of a torus of size $s$ can complete all transmissions in $s\alpha + \beta m$ instead of $s(\alpha + \beta m)$ in the unicast case with dependencies between each flow along each link. 
However, this difference does not affect the completion time for most tori. The CCT of an AllGather on a torus is bound by the diameter of the graph for latency cost and the access bandwidth of each node for bandwidth cost.
The transmission costs along the longest path (which multicast would reduce) are smaller than the access bandwidth cost in most cases as $\frac{k_t}{2} N^{1/k_t} < \frac{N-1}{2k_t}$ for sufficiently large tori  (\eg for $k_t=3$, this is true for $N \geq 64$). 
Similarly to the Clos, an AlltoAll has nothing to gain from multicast as each flow between all pairs of nodes is a unique piece of data.

For an AllReduce, the baseline implementation is an AllGather in each dimension followed by  each node independently reducing the data. 
Thus, this follows the reduction in the longest path time above for a one-dimensional torus (in each dimension). While the diameter does dominate for a AR, the maximum distance of a message ($\frac{k_tN^{1/k_t}}{2}$) is only slightly larger than the access-based bandwidth cost ($\frac{N-1}{2k_t}$) for a 1-D torus ($k_t = 1$). Thus, the saved transmission cost on the longest path does not significantly affect the overall CCT (as it is also bound by the access bandwidth of a node in each dimension).

\takeaway{Multicast only provides a benefit to AllGather (or AllGather-based AllReduces). On both topologies, this only saves extra transmissions / bandwidth cost rather than lowering the best-possible CCT. }
\vspace{-1em}

\section{Case Studies}
\label{sec:case-studies}
We now apply our performance models to determine how each topology affects the communication overheads of training modern LLMs in two specific case studies: the largest model used to benchmark Megatron-LM \cite{megatron-repo, megatron-lm} and DeepSeek v3 \cite{deepseekv3technicalreport}. 
To use our performance models we need both (1) the size (or, degree) of each form of parallelism, and (2) the message size for each parallelism. 
The size is deployment-specific, so we use the values given directly in the benchmark information in \cite{megatron-lm}. 
The message size can then be calculated using model parameters (\eg batch size, sequence size, \etc) \cite{scalingmegatron}. 
We consider four forms of parallelism: Tensor Parallelism (TP), Data Parallelism (DP), Expert Parallelism (EP), and Pipeline Parallelism (PP). As mentioned in \S\ref{sec:background}, these correspond to uses of AllGather, AllReduce, AlltoAll, and point-to-point communication respectively.
We summarize these values for each model in Table~\ref{tab:model-params}.

\begin{table*}
{\small
\begin{tabular}{|c| c|c|c|c|c| c|c|c|c| c|c|c|c|}
    \hline
    \textbf{Model} & \textbf{Params} & \textbf{$s$} & \textbf{$h$} & \textbf{$b$} & \textbf{$l$} & \textbf{TP} & \textbf{PP} & \textbf{DP} & \textbf{EP} & \textbf{$m_{TP}$} & \textbf{$m_{PP}$} & \textbf{$m_{DP}$} & \textbf{$m_{EP}$} \\
    \hline
    Megatron-LM Ex. & 462B & 4096 & 18432 & 1 & 112 & 8 & 16 & 48 & - & 132MB & 18.9MB & 7.2GB & - \\
    \hline
    DeepSeek V3  & 671B & 4096 & 7168 & 1 & 61 & 1 & 16 & 2 & 64 & - & 58.7MB & 1.3GB & 28MB \\
    \hline 
\end{tabular}
}
\caption{Model parameters for our two case studies \cite{megatron-repo,deepseekv3technicalreport} ($s$ is the sequence length, $h$ is the hidden size, and $b$ is the micro-batch size). Message sizes for each parallelism are derived with the other model parameters listed according to \cite{scalingmegatron}. We assume a $b = 1$ for the Megatron-LM example as its value is not provided.}
\label{tab:model-params}
\end{table*}

For the torus, we first assume the typical placement with one parallelism along each dimension (we consider other placements later in this section). Thus, we assume tori of sizes 8x16x48 and 2x16x32.\footnote{The 8x16x48 torus is notably larger than TPUv4 clusters (4096 TPUs) \cite{tpuv4}. We ignore this for now and assume a torus of the correct size is available. Also, in the equations presented in \S\ref{sec:baseline} we assume tori with equal size in each dimension for simplicity. This analysis thus requires a slight change to the equations to generalize to other shapes explained further in Appendix~\ref{app:alt-torus-shapes}.}
Since each parallelism operates in a different dimension, we calculate the communication time for each parallelism on independent slices of the torus (\eg TP between GPUs in the same PP and DP groups will reside in one 1-D torus slice as in the left side of Figure~\ref{fig:placement-visual}). 

For the Clos, we also follow the standard heuristics for placement \cite{topo-opt, scalingmegatron} and place TP and EP (the most data-intensive) to have the highest locality (\eg within a rack), then DP, then PP. 
Such a placement can be seen in Figure~\ref{fig:placement-intro} with P1 = TP, P2 = DP, and P3 = PP.
As before, we assume $k_c=128$, and thus the network is a fixed size regardless of how many nodes are used.

The  CCT (calculated using the equations in Table~\ref{tab:baseline-ub}) for each corresponding collective is shown in Table~\ref{tab:perf-comparison-case-study}.\footnote{To calculate the CCT for pipeline parallelism, we calculate the time for a flow between two corresponding nodes in consecutive stages to complete on the topology. For example, in a torus with PP in one dimension, this will be $\alpha + \beta m$.}
Notably, each collective may not be entirely ``exposed'' \cite{gpipe, 99flops, deepseekv3technicalreport}; \ie there may be computation occurring during communication, meaning the training job is not fully blocked on the network. Thus, these values do not directly represent a per-iteration overhead (which may be smaller if there is overlap), but give a general idea of the time spent completing communication.

As seen in the table, the Clos achieves lower or the same CCTs for most collectives. 
The only exception is for AllGather, which as seen in \S\ref{sec:baseline} performs better on the torus when the two topologies have the same link bandwidth (as the torus has more per-node bandwidth available). 
As mentioned in \S\ref{sec:baseline}, it is possible that a torus network may have higher link bandwidths available than a Clos, given the different link technologies.
Accordingly, we determine how much more per-link BW the torus would need over the Clos to achieve comparable performance in AR with Megatron-LM and ATA with DeepSeek. In both cases, the torus would require about 8x as much per-link bandwidth to achieve similar performance. 

\begin{figure}
    \centering
    \includegraphics[width=0.85\columnwidth]{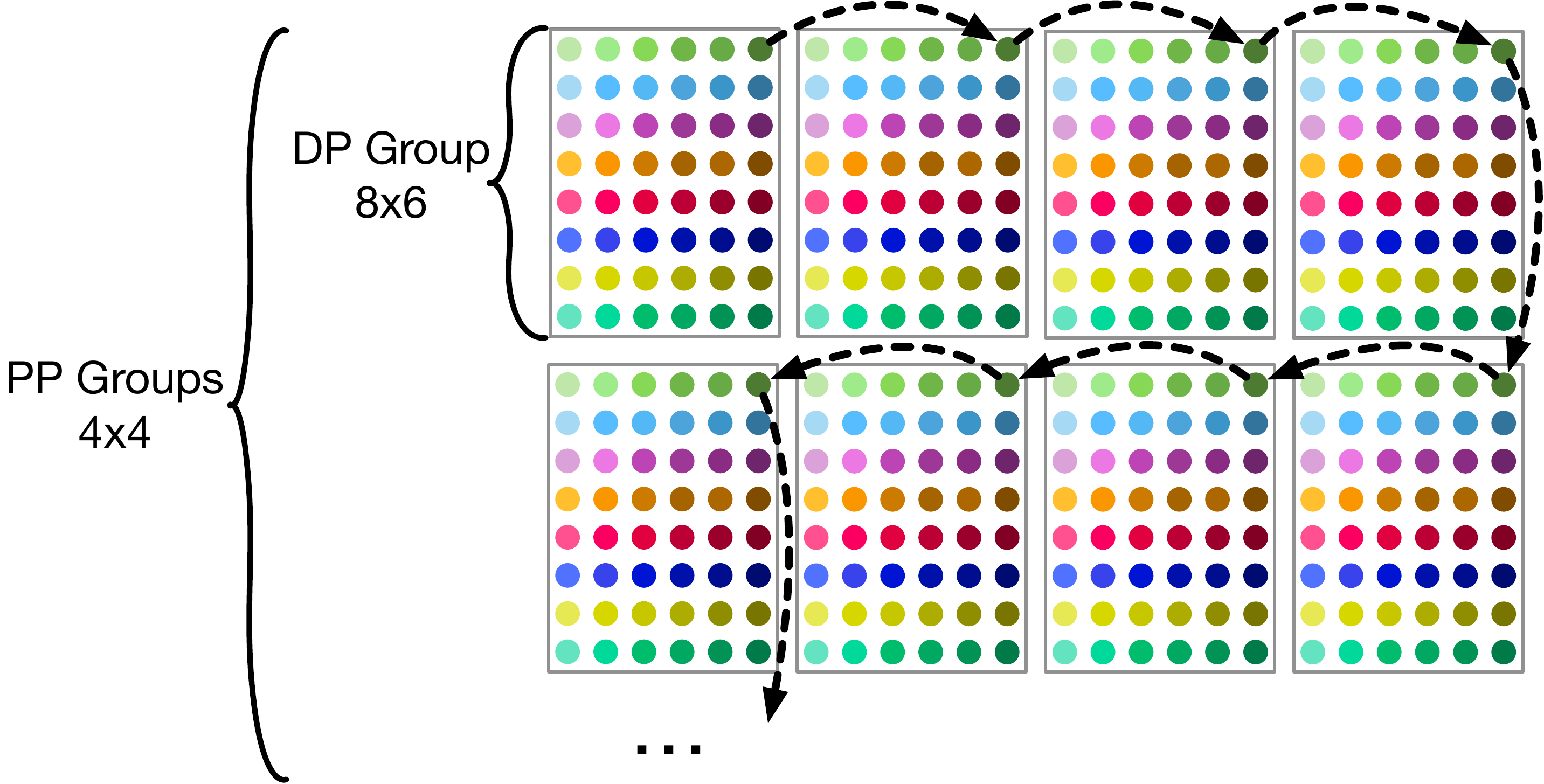}
    \caption{Locality-Opt torus placement for the Megatron-LM case study shown in a 2D slice. DP groups are 8x6 node tiles (denoted with boxes) and PP is performed among nodes of the same color between tiles (16 tiles in a 4x4 grid). The pattern of traffic for the P2P flows between PP peers are shown for one set.  }
    \label{fig:torus-tiling-example}
\end{figure}

As mentioned in \S\ref{sec:placement}, placements optimized for locality in one parallelism can improve the overall performance on the torus in some cases. 
Thus, we consider the Locality-Opt placement strategy.
In the Megatron-LM example, the data parallel groups are quite large and hence have high completion times when embedded in one dimension.
Accordingly, we consider a placement that embeds the DP groups in 2 dimensions of the torus, and then tile these groups in the same dimension with another parallelism.
We choose PP since it has small message sizes and only requires point-to-point communication.
In order to keep locality high, we propose the tiling shown in Figure~\ref{fig:torus-tiling-example} where the 48 degree DP groups are in 8x6 groups which are tiled in a 4x4 grid. The TP groups still reside in the third orthogonal dimension (requiring a 8x32x24 torus instead of the previous 8x16x48). 
Importantly, these tiles do not have wraparound links, so the locality is not as good as an 8x6 torus (where the max distance between nodes is halved).
Between each DP group, nodes will communicate with the subsequent tile to form the PP groups (corresponding colors in the figure). In the worst-case, this requires PP traffic to travel 8 hops and compete for bandwidth with 8 other flows (when traversing down). 

With this placement, the AG (TP) performance remains the same, the AR (DP) improves with the better locality between nodes, and the P2P between PP nodes gets worse as the distance between nodes (and demand on a given link) increases. The exact CCTs are shown in Table~\ref{tab:perf-comparison-case-study}. 
We apply a similar placement to the EP groups in the DeepSeek example, again with a 4x4 grid of higher-locality groups. In this case, however, the EP groups are the tiles and are of size 8x8. This placement gives the CCTs shown in the table. Again, the EP group improves its CCT, but performs worse in PP. Notably, these improvements are limited due to the lack of wraparound links for a given tile.

Lastly, we consider a hybrid topology: tori connected via a Clos (as discussed in \S\ref{sec:hybrid}).  
We assume the specific topology of TPUv4 \cite{tpuv4} where each rack in this Clos is 64 nodes in a 4x4x4 block and a rack is 64 of these blocks.
As shown in the table, the hybrid topology can leverage the additional connectivity to achieve generally the best performance of the other options. The only exception is the AG in DeepSeek, which requires traversing racks via the torus and thus incurs additional delay. 

\takeaway{In our case studies, the Clos outperforms the torus in ATA and AR, though the torus has lower CCT for AG. The torus's performance can be improved in some collectives with locality-optimized placements (\ie allowing a parallelism to use more than one dimension), but the benefits are limited by the lack of wraparound links. A torus-Clos hybrid outperforms in all cases.}

\begin{table}
{
\small
\vspace{-1em}
\begin{tabular}{|c| c|c|c|c|c|}
    \hline
    \textbf{Model} & \textbf{Topology} & \textbf{AR} & \textbf{AG} & \textbf{ATA} & \textbf{P2P} \\
    \hline
    Megatron-LM & Torus & 4.83 s & 9 ms & - & 0.4 ms \\
    \cline{2-6} 
    Example & Clos & 1.12 s  & 18 ms & - & 0.4 ms   \\
    \cline{2-6} & Torus LC & 4.03 s  & 9 ms & - & 3 ms   \\
    \cline{2-6} & Hybrid & 1 ms  & 9 ms & - & 0.4 ms   \\
    \hline
    \multirow{2}{*}{DeepSeek v3 } & Torus & - & 36 ms & 282 ms & 26 ms \\
    \cline{2-6}
    & Clos & - & 36 ms & 35 ms & 26 ms\\
    \cline{2-6} & Torus LC & - & 36 ms & 184 ms & 78 ms   \\
    \cline{2-6} & Hybrid & - & 114 ms & 17.6 ms & 26 ms   \\
    \hline 
\end{tabular}
}
\caption{Best-case CCT for case study models (Table~\ref{tab:model-params}) for each collective. The default values for $\alpha$, $\beta$, \etc from Table~\ref{tab:variables} are used. Torus LC denotes a torus placement with locality optimized for the slowest parallelism/collective. }
\label{tab:perf-comparison-case-study}
\vspace{-1.5em}
\end{table}

\section{Discussion}

We started this paper thinking that an analytical approach would provide clear and rigorous answers to which topology should be the basis for serving ML training workloads. 
What we found out, after looking at the problem from many different perspectives (from developing a baseline model, to considering failures, placement, cost, and hybrid designs) and exploring the impact of other factors (such as message sizes and multicast functionality), is that the answer of which topology is better is, quite bluntly, an unsatisfying ``it depends.'' However, there are a few general trends that we can call out: 
(i) For AllGather, the torus generally outperforms the Clos;
(ii) as the number of nodes increases, the Clos tends to outperform; 
(iii) under realistic placement and torus dimensionality, the torus \emph{only} outperforms the Clos on AllGather (and does so regardless of the number of nodes); and 
(iv) the performance on the torus is far more sensitive to link failure and placement than on the Clos.

If forced to provide a one-sentence description of our findings, we would say that unless AllGather is your bottleneck for performance, the Clos topology is likely the better choice.

\bibliographystyle{ACM-Reference-Format} 
\bibliography{sections/biblio}

\clearpage
\appendix{
\section{Failure Derivations}
\label{app:failure-derivations}

In this section, we derive the results shown in Table~\ref{tab:failure}.
As in the table, we will use $R=k_c/2$ and $P=k_c^2/4$ for the number of nodes in a rack and a pod in a Clos respectively.

\subsection{Network-Level Mitigations}
In network-level mitigations, the algorithm used to implement the collective, or the ``schedule'' is unchanged. Thus, the same series of source-destination pairs will communicate. Only the routing or load-balancing in the network may change to avoid failures.

\subsubsection{AllReduce}

\paragraph{Clos} 
In the baseline algorithm, nodes exchange their messages in pairs and reduce before exchanging with a different pair (recursive doubling \cite{topologyaware-sack12}). This pattern occurs stages: within a rack, within a pod, and between pods (this specific order is not important though).  

First, we consider an L1 failure (between a ToR switch and an aggregation switch).
In this case, under the existing algorithm, the first $log_2(R)$ transmissions for each node will proceed as normal since all host-to-ToR links are up. 
All subsequent steps of the algorithm will be affected by the failures as each flow uses one path of all the available paths within the pod or between pods.
Thus, in $log_2(N)-log_2(R) = log(\frac{N}{R})$ rounds, an additional $\beta m$ must be incurred due to the flow displaced by the failure using the same link as another -- \ie the two flows must share the link and thus incur additional bandwidth cost. 
No additional latency cost is necessary as the path is the same length as the failed alternative.
For all rounds after the initial within-rack stage, this gives $log_2(\frac{N}{R})\beta m$.

The derivation for an L2 failure is the same, except that only $log_2(\frac{N}{P})$ stages are affected. 

\paragraph{Torus} 
In the baseline algorithm, each node participates in an AllReduce in each dimension of the torus sequentially.
Thus, for a single-link failure, only one stage (an AllReduce in one dimension) of the algorithm will be affected.
We assume these AllReduces are implemented as an AllGather followed by a reduction since this minimizes the necessary distance messages must go around the ring (vs. reducing as a single message traverses the ring).
Thus, the completion of this AR is bound by the diameter of the ring $\frac{N^{1/k_t}}{2}$ and each transmission and propagation delay along it.
Reductions must be done in each ring at the end, and can be done in $log_2(\frac{N^{1/k_t}}{2})$ operations.

Each of the messages on a failed link will have to be re-routed through another dimension of the torus (using a 3-hop route as in Figure~\ref{fig:torus-failure}).
These additional hops will add $2\alpha$ latency to the path of all messages that that traverse it.
Similarly, the flow will compete for bandwidth on the second link in the 3-hop detour as an AllGather is occuring in all other rings in the same dimension in the torus at the same time.
Thus, an additional $\beta m$ must be paid as well.
In total, this gives an additional $2\alpha + \beta m$.

\subsubsection{AllGather}

\paragraph{Clos} 
In the Clos, the baseline AllGather algorithm simply has each node send a copy of its message to all other nodes. In the no-failure case, both sending the messages iteratively and in parallel achieves the minimum CCT. In the failure case, this distinction does matter as it affects the demand on L1 and L2 links. In the iterative version, L1 links, for example, are unused for the first $R$ transmissions, then are used by one node for all subsequent transmissions. In the parallel case, the nodes can only send as fast as their access link and thus L1 and L2 links are not fully utilized since some flows do not need to traverse them. Thus, we consider the parallel case for failure as there is more free capacity for displaced flows.

When there is a L1 failure, flows starting in rack with the failure destined to any node in a different rack are affected. We only consider the upward direction (ToR to aggregation) without loss of generality (as long as upward paths are balanced correctly across available paths).
For a given ToR with the L1 failure, it will have $R=k_c/2$ nodes below it sending on the $k_c/2 - 1 = R-1$ links with the one failure. 
Thus, $R(N-R)$ flows must traverse these $R-1$ links (since each node will send to $N-R$ other nodes via these links).
The bandwidth cost associated with this is then $\lceil \frac{R(N-R)}{R-1}\rceil \beta m$. Note that flows cannot be infinitely divided (unless we assume per-packet load balancing), hence the ceiling.
Subtracting the baseline bandwidth cost gives $\left( \left\lceil \frac{R(N-R)}{R-1}\right\rceil - (N-1) \right)\beta m $

An L2 failure follows the same derivation except that $P$ nodes send $N-P$ messages on $P-1$ links.

\paragraph{Torus} 
In the baseline AllGather algorithm on a torus, each node initially sends its message in all directions. After this, nodes propagate each message they receive to neighboring nodes that have not received the message. We do not prescribe the exact algorithm as many approaches to propagation perform equivalently. 
Thus, the CCT is bound both by the diameter of the network and the access bandwidth of the nodes. However, the access bandwidth almost always dominates in tori with more than 1 dimension (\ie $\frac{k_t N^{1/k}}{2} < \frac{N-1}{2k}$ for $k > 1$ for most values of $N$).

Under a link failure, since all flows are single-hop in the schedule, a detour into another dimension will be required again. 
Thus, $2\alpha$ will be added to the latency cost.
While the detour will also incur a bandwidth cost as the flow competes with other flows, this will only affect the time required for the longest path, which is not a dominant factor.
Instead, the more important factor is that the nodes adjacent to the failure will not be able to receive ${N-1}{2k_t}$ of their messages on the failed links.
Accordingly, these messages will be routed onto a different access link, doubling its associated bandwidth cost and incurring $\frac{N-1}{2k_t}\beta m$.
Altogether, this gives an additional $2\alpha + \frac{N-1}{2k_t}\beta m$

\subsubsection{AlltoAll}

\paragraph{Clos} 
The derivation for AlltoAll on the Clos exactly follows the AllGather derivation since their schedules are the same (send one message to all other nodes iteratively or in parallel).

\paragraph{Torus} 
In the baseline schedule on the torus, each node simply sends to all other nodes on their shortest paths.
We do not prescribe which shortest paths as many are possible and achieve the same CCT.

Deriving the exact impact on CCT would likely require determining which of the many potential alternate shortest paths each flow that traverses a particular failed link should take. Since this would require prescribing an initial set of paths, we instead derive a lower bound.

To determine the lower bound, we take our original derivation and change the number of available links for all necessary capacity to traverse. 
The completion time without failures is determined by the total capacity required in the network and the number of links.
Since the average distance between any pair of nodes is $\frac{k_t}{4}N^{1/k_t}$ and the total number of messages to be sent is $N(N-1)$, the total required capacity is $\frac{k_t}{4}N^{1/k}N(N-1)$.
The total number of directed edges in the topology is $2k_tN$.
Thus, the minimum CCT is bound by $\frac{(N-1)N^{1+\frac{1}{k_t}}k_t}{8kN}$.

With a single link failure, we determine the lower bound by dividing by the total number of links \emph{with} the bidirectional failure.
Thus the new potential completion time is $\frac{N(N-1)\frac{k}{4}N^{1/k}}{2kN - 2}$. 
To get the \emph{additional bandwidth cost}, we subtract the baseline: $\frac{N(N-1)\frac{k}{4}N^{1/k}}{2kN-2} - \frac{(N-1)N^{1/k}}{8}$.
This gives $\frac{(N-1)N^{1/k}(k-1)}{8(kN-1)} $

\subsection{Schedule-Level Mitigations}
In schedule-level mitigations, the algorithm used to implement the collective, or the ``schedule'' \emph{can} be modified to avoid the failure.

\subsubsection{AllReduce}

\paragraph{Clos} 
In order to minimize the impact of a failure, the schedule can be modified assume one less node is participating for most of the collective.

More specifically, for either L1 or L2 failures, the within-rack stage of an AllReduce can proceed as normal. Once this is complete, a failure will affect the collective. 
However, this can be avoided by simply assuming that one node under the ToR with the failure is not participating in the collective. 
The baseline algorithm can then proceed as usual with $N-1$ nodes. Since the under-ToR stage has already occurred the data from the excluded node will be included in future exchanges with other nodes by nodes in its rack. 

At the end of the collective with $N-1$ nodes, the last node must receive the final reduced message with all other node's data, incurring $2\alpha + \beta m$ if received from a node in its rack.

\paragraph{Torus} 
In a torus, the relatively small overhead of a failure for AllReduce can be reduced slightly.
The algorithm should perform the per-dimension AllReduces in each dimension that the failed link is not in. Then, when performing the AR in the failed dimension, the ring with the failed link does nothing.
Finally, the nodes in the ring of the failed link receive the final reduced message from some other adjacent node in a different ring. 
This final, extra transmission requires an additional $\alpha + \beta m$.

Notably, we assume a $k_t>1$ as in a ring there are no other dimensions to exploit and the CCT must reflect the $N-1$ diameter of the network.

\subsubsection{AllGather}

\paragraph{Clos} 
The baseline AllGather algorithm has no dependencies between flows, and thus nothing to change while maintaining the same CCT (adding a dependency will in the very least add more latency). 

Accordingly, we consider other algorithms which do introduce dependencies and thus may be worse than network-level approaches.
Following the structure of the AllReduce approach, the schedule could ignore one node that would have used the failed link and then have another node in its rack propagate the messages. However, because there are not reductions in this case, this incurs both a latency cost and a significant bandwidth cost: $2\alpha + (N-R)\beta m$ as the other node has to send $N-1$ messages of $m$ size.

Assuming that bandwidth/message size is more dominant than latency, the better approach is to use an algorithm like a ring algorithm. In this case, each node sends to only one other node for the entire collective. As in a torus, each node propagates each message it receives to its ``neighbor''. With such a schedule, few links in the network must be used.
For example, a ring could have mostly intra-rack flows and only cross the boundaries of a particular rack/pod twice (in and out). 
Thus, a failure could be easily avoided at the L1 or L2 level since most links at these levels are not used.
However, the dependencies increase the latency significantly and incur a latency cost of $\frac{N-1}{2}\alpha$, based on the diameter of the ring. 
Notably, while this uses host access links to send to two neighbors to minimize the longest path of a message, this does not affect the bandwidth cost as each flow takes $2\beta m$, but only half the transmission rounds are required.

\paragraph{Torus} 
As with the ATA with network-level mitigation on torus, we determine a lower bound for the CCT rather than prescribe a particular series of flows to propagate each message and determine alternate paths.
The bandwidth cost must increase due to the lost access bandwidth of the nodes adjacent to the failure. 
Thus, the bandwidth cost with failure is $\frac{N-1}{2k_t -1}\beta m$. 
Subtracting the baseline of $\frac{N-1}{2k_t}\beta m$ gives $\frac{N-1}{2k_t(2k_t -1)}\beta m$.

\subsubsection{AlltoAll}

There are no schedule-level mitigations with an AlltoAll since there are no dependencies in the schedules. Adding dependencies may help avert a failure, but would require sending data to a node that does not need it. We do not consider this case.

\section{Placement Derivations}
\label{app:placement-deriv}

In this section, we derive the results shown in Table~\ref{tab:placements}. 
These results assume a standard $2$-dimensional torus as the base network, with $\sqrt{N}$-sized parallelized groups.

\subsection{Typical Placement}

As shown in Figure~\ref{fig:placement-visual}, the typical placement of collectives in a $2$-D torus places each group along one axis. 
The set of links that directly connect nodes participating in the same collective forms a ring of size $\sqrt{N}$.
The collective completion times for AllReduce, AllGather, and AlltoAll can therefore be computed from the values in Table~\ref{tab:baseline-ub} plugging in values $k_t = 1$ and $n = \sqrt{N}$.
Therefore, we can derive the sum cost of AlltoAll plus AllReduce as
\begin{align*}
& \alpha \frac{\sqrt{N}}{2} + \beta m \sqrt{N} \cdot \frac{\sqrt{N}}{8} + \frac{\sqrt{N}}{2}(\alpha + \beta m) + \gamma m \log_2(\sqrt{N}) \\ 
& = \alpha \sqrt{N} + \beta m \left(\frac{N}{8} + \frac{\sqrt{N}}{2}\right) + \frac{1}{2}\gamma m \log_2\left(N\right) .
\end{align*}
Meanwhile, the sum cost of AlltoAll plus AllGather is
\begin{align*}
    & \alpha \frac{\sqrt{N}}{2} + \beta m \sqrt{N} \cdot \frac{\sqrt{N}}{8} + \frac{\sqrt{N}}{2}(\alpha + \beta m) \\
    & = \alpha \sqrt{N} + \beta m \left(\frac{N}{8} + \frac{\sqrt{N}}{2}\right) .
\end{align*}

\subsection{Locality-Opt Placement}

As shown in Figure~\ref{fig:placement-visual}, the Locality-Opt strategy optimizes for locality of the parallelized collective that is most bandwidth-intensive. 
In our case, we assume this to be AlltoAll.
Thus in our derivation, the AlltoAll collective groups tile the torus into $\sqrt{N}$ grids, each of size $N^{1/4}$ by $N^{1/4}$.
For the secondary collective, either AllReduce or AllGather, groups are made up of representatives from each tile (e.g. one group is made up of all nodes in the top right of their respective tile).
This placement of groups depending on the collective is made below when examining the derivations for each collective completion time.
Entries in Table~\ref{tab:placements} may be computed by summing the appropriate derivations.

\subsubsection{AllReduce}

AllReduce latency cost remains $\alpha$ times the diameter of the group, as always. 
However, since each group is made up of one representative from each tile, the diameter of each group may be up to $2 \sqrt{N} = 2 \sqrt{N}$. 

AllReduce bandwidth cost is at least $\beta m$, because each node must at least receive a single reduced message.
We can implement the AllReduce by first reducing between group members along the same row, then reducing between groups members along the same column. 
This splits the AllReduce into two stages: first, nodes receive $N^{1/4}$ messages and reduce them together. Then, nodes receive another $N^{1/4}$ messages and reduce those together. Note that receiving $N^{1/4}$ message along a single axis of the torus takes $\frac{1}{2} \beta m N^{1/4}$ bandwidth cost, because there are $2$ incoming links per axis that may receive messages.
In total, this would take $\beta m N^{1/4}$ bandwidth cost, and $2 \gamma m \log_2\left(N^{1/4}\right)$ computation cost.

This allows us to derive a total completion time upper bound of 
\[
    2 \alpha \sqrt{N} + \beta m N^{1/4} + \frac{1}{2} \gamma m \log_2(N) .
\]

\subsubsection{AllGather}

The latency cost of AllGather is the same as AllReduce, $2 \alpha \sqrt{N}$. 
The AllGather bandwidth cost is $\beta m$ times the number of messages that need to be received by each node (which is the size of each collective group, $\sqrt{N}$) divided by the total number of incoming links per node, $4$.
Therefore, the total completion time can be bound by 
\[ 
2 \alpha\sqrt{N} + \frac{1}{4}\beta m \sqrt{N} . 
\]

\subsubsection{AlltoAll}

The latency cost of AlltoAll, like the above collectives, is $\alpha$ times the diameter of the group.
However, since each group is a tile of size $N^{1/4}$ by $N^{1/4}$, the diameter is no more than $2N^{1/4}$. 
Thus, the overall latency cost is $2 \alpha N^{1/4}$.

The bandwidth cost of AlltoAll is more complex.
It is $\beta$ times the maximum amount of flow that must traverse any single link. 
Because we assume optimal routing that perfectly load-balances links, this is equal to the average amount of flow traversing links.
(Note that because many flows traverse many links, this is greater than the number of total messages.)
The average amount of flow is the number of total source-destination pairs in a single collective group ($\sqrt{N}^2$) times the average path length between pairs ($\leq N^{1/4}$), divided by the total number of links used by the collective group (about $4\sqrt{N}$).
Thus, we can derive the total completion time as no more than
\begin{align*}
   & 2 \alpha N^{1/4} + \beta m \frac{N \cdot N^{1/4}}{4\sqrt{N}} \\
   & = 2\alpha N^{1/4} + \frac{1}{4} \beta m N^{3/4} .
\end{align*}

\section{Multicast Derivation}
\label{app:multicast}

In this appendix, we briefly derive the transmissions saved with multicast in an AllGather on a Clos.
In the unicast case, each node will send a message to all other nodes.
For a destination in the same rack, the data will be transmitted on 2 hops.
For a destination in the same pod (but not the same rack), this is 4 hops and all others are 6.
There are $k_c/2$ nodes in a rack, $k_c^2/4$ in a pod, and $k_c^3/4$ in total.
Thus, the unicast baseline will send a message across $2(k_c/2 -1) + 4(k_c^2/4 - k_c/2) + 6(k_c^3/4 -k_c^2/4)$ links for \emph{each} node.
Overall, there are $(k_c^3/4) \left[ 2(k_c/2 -1) + 4(k_c^2/4 - k_c/2) + 6(k_c^3/4 -k_c^2/4) \right]$ transmissions of messages.

In the multicast case, each node only has to send its message on a spanning tree to all other nodes in the network. This will require all access links ($k_c^3/4$), a link to each other ToR ($k_c^2/2 - 1$), a link to each other pod ($k_c -1$), and a route up to each level of the network (+3). 
Altogether, this gives $k_c^3/4 + k_c^2/2 - 1 + k_c -1 + 3$ transmissions per node and $(k_c^3/4) \left[ k_c^3/4 + k_c^2/2 + k_c + 1 \right] $ overall.

The difference between the unicast count and multicast count is $ \frac{k^3}{16} (5 k^3 - 4 k^2 - 8 k - 12)$
\section{Hybrid Topology Derivations}
\label{app:hybrid-deriv}
In this section, we provide the derivations for all results in Table~\ref{tab:hybrid}. While the exact forms of the equations generally come from the results in \S\ref{sec:baseline}, we explain the algorithms behind them here. 
Notably, many algorithms are possible and we do not claim these are optimal. Instead, we aim to create well-performing options following the heuristic of ``use the scale-up topology whenever possible.''

\subsection{Clos + Mesh}
\subsubsection{AllReduce}
The overall algorithm remains fairly similar to the baseline Clos in this case. Since the mesh only spans a rack, it is used for the intra-rack stage of the AllReduce. All nodes in the rack exchange messages and perform reductions. Then, the remainder of the algorithm proceeds as it does on only a Clos: nodes exchange their data in pairs.
Thus, the CTT is the sum of these two phases.
The mesh portion only requires $\alpha + \beta m + \gamma m R$ as each GPU is directly connected to all others. Therefore, a node can transmit to all others in the rack in parallel, and then perform the reduction.
The remainder of the algorithm then follows the baseline Clos, but for $log_2(N)-log_2(R)$ exchanges.

\subsubsection{AllGather}
An AllGather can similarly algorithmically exploit the faster connectivity of the mesh. Nodes can exchange messages with nodes in their rack and in the rest of the network in parallel.
The intra-rack exchange will only take $\alpha + \beta_u m$ while the Clos-based exchange will inevitably take longer.
However, nodes do not need to send a message to all other nodes via the Clos: they can send one copy to each rack and use the mesh to quickly replicate it.
Therefore, the scale-out network phase takes $ 6\alpha + (\frac{k_c^2}{2}-1) \beta_o m $ where  $\frac{k_c^2}{2}$ is the total number of racks in the network.
Together with the final intra-rack exchange where each node sends $\frac{k_c^2}{2}-1$ messages to each other node we have: $ 6\alpha + (\frac{k_c^2}{2}-1) \beta_o m  + \beta_u m (\frac{k_c^2}{2}-1)$

\subsubsection{AlltoAll}
In the AllToAll, there are no algorithmic changes to make without introducing unnecessary dependencies. Thus, the scale-up network can be used for intra-rack connections while the scale-up is used for all remaining.
Therefore, the CCT will be $max(\alpha + \beta_um, 6\alpha+ \beta_om(N-R)$ which is trivially $6\alpha+ \beta_om(N-R)$.

\subsection{Clos + Torus}
\subsubsection{AllReduce}
Similarly to the mesh, our algorithm performs an AllReduce within the torus (following the approach from \S\ref{sec:baseline}) before completing it via the Clos.
Accordingly, the first phase takes $\frac{k_tP^{1/k}}{2}(\alpha + \beta_um + \gamma m$ as the pod of size P performs the AR.
Then, the normal AR algorithm on Clos completes the collective with $log_2(N)-log_2(P)$ exchanges.

\subsubsection{AllGather}
The AllGather also follows a similar structure to the above. First, an AllGather is performed within each torus while nodes distribute their messages to \emph{one} node in all other pods. This will take $max(\frac{k_t P^{1/k_t}}{2}\alpha +  \beta_u m \frac{P-1}{2k_t}, 6\alpha+ \beta_om(k_c-1))$. With our default values for all variables and our assumed scale ($k_c=128$ and therefore $P=4096$) the torus portion actually takes longer (about 170ms vs 32ms). However, we assume $\beta_u > \beta_m$ by a large enough ratio that the torus is in fact faster (>5x). 

Lastly, once this phase has completed, the torus performs one final AllGather with the new data which takes $\frac{k_t P^{1/k_t}}{2}\alpha +  \beta_u m (k_c -1)\frac{P-1}{2k_t}$ since each node has $k_c-1$ new messages to exchange.
Altogether, this gives $6\alpha+ \beta_om(k_c-1) + \frac{k_t P^{1/k_t}}{2}\alpha +  \beta_u m (k_c -1)\frac{P-1}{2k_t}$.

\subsubsection{AlltoAll}
Again, in an AllToAll, we make no algorithmic changes and simply use the scale-up network when possible with the existing traffic matrix. 
Thus, each tori will perform an AllReduce while the Clos handles all inter-pod connections. 
The torus ATA will take $\frac{k_tP^{1/k_t}}{2}\alpha + \beta_um(P-1)\frac{P^{1/k_t}}{8}$ while the Clos will take $6\alpha + \beta_o m (N-P)$. 
Since the Clos is serving $k_c = 128$ times as many flows as the torus, the max is always the Clos portion, even with $\beta_o=\beta_u$.
Thus, the CCT is $6\alpha + \beta_o m (N-P)$.
\section{Alternate Torus Shapes}
\label{app:alt-torus-shapes}

In the baselines presented in \S\ref{sec:baseline}, we assume a torus of equal size in each dimension. Here, we relax this assumption for use in our case studies. 

In general, this change will only impact the diameter of the network since the the per-node bandwidth remains the same.
Thus, all terms with $\frac{k_t N^{1/k_t}}{2}$ become $\frac{1}{2}\sum_{i=1}^{k_t} s_i$ for a side length, $s_i$ in each dimension $i$. 
}

\end{document}